\documentclass[12pt]{iopart}
\usepackage{xcolor}
\usepackage{amssymb}
\usepackage{array}
\usepackage{graphicx}
\usepackage{hyperref}
\expandafter\let\csname equation*\endcsname\relax
\expandafter\let\csname endequation*\endcsname\relax
\usepackage{amsmath}

\newcommand{\Tc}{T_\mathrm{c}}

\begin{document}

\title{Evidence of de Almeida--Thouless line below six dimensions}

\author{M.~Aguilar-Janita}
  \address{Departamento de F\'{\i}sica,
  Universidad de Extremadura, 06006 Badajoz,
  Spain}
\author{V.~Martin-Mayor}\address{Departamento de F\'\i{}sica
  Te\'orica, Universidad Complutense, 28040 Madrid,
  Spain}
  
\author{J.~Moreno-Gordo}\address{Departamento de Didáctica de las Ciencias Experimentales y de las Matemáticas, Universidad de Extremadura, 10004 Cáceres, Spain}\address{Instituto de Computaci\'on Cient\'{\i}fica Avanzada (ICCAEx), Universidad de Extremadura, 06006 Badajoz, Spain}\address{Instituto de Biocomputaci\'on y F\'{\i}sica de Sistemas Complejos (BIFI), 50018 Zaragoza, Spain}

\author{J.J.~Ruiz-Lorenzo}\address{Departamento de F\'{\i}sica,
  Universidad de Extremadura, 06006 Badajoz,
  Spain}\address{Instituto de Computaci\'on Cient\'{\i}fica
  Avanzada (ICCAEx), Universidad de Extremadura, 06006 Badajoz,
  Spain}

\vspace{10pt}
\begin{indented}
\item[] \today 
\end{indented}

\begin{abstract}
We study the critical behavior of the Ising spin glass in five spatial dimensions through large-scale Monte Carlo simulations and finite-size scaling analysis.
Numerical evidence for a phase transition is found both with and without an externally applied magnetic field.  The critical exponents  are computed in both cases. We compute with a 10\% accuracy the lower critical dimension at zero magnetic field, finding a result consistent with estimates obtained with entirely different methods, by combining our estimates of critical exponents in five dimensions with previous results for other spatial dimensions.  When the results in a magnetic field are compared with previous results in six spatial dimensions, qualitative differences emerge in the scaling behavior of the correlation functions at zero external momentum. This anomalous scaling does not extend to other wavevectors.  We do not find indications of a quasi first-order phase transition in a magnetic field. 
\end{abstract}

%

\submitto{\JSTAT}

\maketitle
%
%
\section{Introduction}
Some of the most basic notions in the theory of critical phenomena and the renormalization group (RG)~\cite{wilson:74,parisi:88,amit:05} become suspect when the magnetically disordered alloys known as spin glasses~\cite{mydosh:93,young:98,charbonneau2023spin} are subjected to an externally applied magnetic field. We are referring, in particular, to the lower $d_\text{l}$ and the upper critical dimension $d_\text{u}$ ---no phase transition appears if the space dimension is $d<d_\text{l}$, while Mean Field approximation provides correct predictions for the critical exponents if $d>d_\text{u}$. Indeed, the value of $d_\text{l}$ and $d_\text{u}$ still remains controversial in spite of the significant uninterrupted efforts of the community~\cite{bray:80,pimentel:02,mcmillan:84,fisher:86,bray:87,fisher:88,yeo:15,bray:11,parisi:12,singh:17b,charbonneau:17,charbonneau:19,holler:20,janus:12,janus:14b,janus:14c,katzgraber:05b,katzgraber:09,vedula:23,vedula:24,vedula:25,leuzzi:09,dilucca:20,Angelini:22,angelini:25,aguilar:24b}. This confusing scenario is somewhat surprising given the extraordinary success of the RG in the study of disordered systems, including classical~\cite{harris:76,janus:13} and quantum~\cite{miyazaki:13,bernaschi:24b} spin glasses in zero magnetic field.

Mean field theory predicts that, even in a magnetic field, a spin-glass transition will occur upon lowering the temperature at the so called de Almeida-Thouless (dAT) line~\cite{dealmeida:78}. Thus, the next logical step would be identifying $d_\text{u}$ and computing the critical exponents. Indeed,  
replicated field-theory predicts $d_\text{u}=6$, but the failure to
find, in the one-loop approximation, a fixed point stable at $d=6$~\cite{bray:80,pimentel:02} perplexed researchers. Some alternatives have been suggested to exit the impasse. For instance, some droplets model supporters think that $d_\text{l}=\infty$ (i.e. the dAT line would not exist in any finite space dimension)~\cite{mcmillan:84,fisher:86,bray:87,fisher:88}. However, a softer version of this claim has been advanced, namely $d_\text{l}=d_\text{u}=6$ which implies that a dAT line would exist only for $d>6$~\cite{yeo:15,bray:11} (mean field theory would be exact at these high dimensions). Both statements have been criticized in Refs.~\cite{parisi:12,holler:20} (in particular, H\"oller and Read suggests that the difficulties are caused by a quasi-first order transition taking place in $d=6$ in a field~\cite{holler:20}). Other authors, using a newly developed loop expansion around the Bethe lattice, have found that $d_{\text{u}}=8$~\cite{Angelini:22} and have even computed critical exponents at first order in a $d-d_\text{u}$ expansion~\cite{angelini:25}. Interestingly enough, a \emph{traditional} two-loop computation has identified a stable, not-Gaussian, RG fixed-point in $d=6$~\cite{charbonneau:17,charbonneau:19} which, qualitatively speaking at least, agrees with the
$d_\text{u}>6$ scenario~\cite{Angelini:22}.

Non perturbative methods, such as Monte Carlo simulations, could be very useful to shed light although, as it has been recently emphasized~\cite{vedula:23,vedula:24,vedula:25}, they suffer from other limitations such as the limited range of system sizes that can be equilibrated. The quasi-first order phase transition scenario has been considered  recently~\cite{fernandez:22,aguilar:24b}, without finding any evidence supporting it. The $d_\text{l}=\infty$ seems to have been ruled out as well. Indeed, numerical evidence for a dAT line in $d=6$ has been found~\cite{aguilar:24b}. Also, a high-temperature series finds a dAT line for $d\ge 6$~\cite{singh:17b}. The finding of a dAT line exactly at $d=6$ (but \emph{not} for $d>6$) has been disputed by Kac-model based studies~\cite{katzgraber:05b,katzgraber:09,vedula:23,vedula:24,vedula:25} ---the Kac-model is a one-dimensional model with long range interactions that mimics the behavior of systems  with short-range interactions in $d$ spatial dimensions--- but other studies claim in favor of a dAT line for much lower $d$~\cite{leuzzi:09,dilucca:20}. According to Refs.~\cite{vedula:23,vedula:24}, the finding of a dAT line at $d=6$ would have been an artifact caused by the small size of the systems that could be brought to thermal equilibrium, but the existence of a dAT line for $d>6$ is undisputed. 

Below $d=6$ there is a smaller number of available results. 
The Janus collaboration, using the custom built computer Janus 1, was able to identify a dAT line 
in $d=4$ and to compute the critical exponents~\cite{janus:12}. However, similar simulations in 
$d=3$ were inconclusive~\cite{janus:14b,janus:14c} 
which suggests that, $3<d_\text{l}<4$ in a magnetic field.\footnote{In the absence of a magnetic field $d_\text{l}\approx 2.5$~\cite{franz:94,boettcher:05,maiorano:18}.} As far as 
we know, the gap between space dimensions $d=4$ and $d=6$ has never been bridged. That is the 
problem that we address in this work.

The remaining part of this work is organized as follows. Definitions for the model and the observables that we compute are given in Sect.~\ref{sect:Model_and_observables}. Our finite-size scaling analysis is presented in Sect.~\ref{sec:FSS} (additional and more technical details will be found in~\ref{app:FSS}).
Our simulations are described in Sect.~\ref{sect:simulations}. The results in zero magnetic field are presented in Sect.~\ref{sect:zero-h}, while the results in a field wild be found in Sect.~\ref{sect:not-zero-h}. Our conclusions are given in Sect.~\ref{sect:conclusions}.

\section{The model and Observables}\label{sect:Model_and_observables}
We perform numerical  simulations of the Edwards-Anderson model for Ising spins (i.e. $s_{\boldsymbol{x}} = \pm 1$), in the presence of an external magnetic field $h$. These simulations are performed on a five-dimensional hypercubic lattice of volume $V = L^5$ with periodic boundary conditions and nearest-neighbor interactions. The Hamiltonian governing the model is expressed as:
\begin{equation}\label{hamiltonian}
    \mathcal{H} = - \sum_{\langle \boldsymbol{x}, \boldsymbol{y}\rangle} J_{\boldsymbol{x} \boldsymbol{y}} s_{\boldsymbol{x}} s_{\boldsymbol{y}}
    -h\sum_{\boldsymbol{x}} s_{\boldsymbol{x}}\;,
\end{equation}
where the coupling constants $J_{\boldsymbol{x} \boldsymbol{y}}$ are independent and identically distributed random variables, taking values of $\pm 1$ with equal probability. Each specific configuration of couplings is referred to as a \textit{sample}. We name \text{replicas} to distinct spin-system copies $\{ s^{a}_{\boldsymbol{x}}\}$ that evolve with identical couplings $\{ J_{\boldsymbol{x} \boldsymbol{y}}\}$  but under statistically independent thermal noise. The superscript $a$ will label the particular replica that we shall be referring to.

Hereafter, we will use $\overline{(\cdots)}$ to denote the average over the couplings, while $\langle(\cdots)\rangle$ will represents the thermal average computed for fixed couplings ${J_{\boldsymbol{x} \boldsymbol{y}}}$.

\subsection{Observables}
In this work we use a parallel approach in the study of the relevant observables of the system as conducted in Ref. ~\cite{aguilar:24b}. Consequently, our focus will be on addressing three specific questions: are we deep enough in the spin-glass phase to observe its characteristic phenomenology? Does the second-moment correlation length exhibit behavior indicative of a phase transition? Do the observables computed to study the RS effective Hamiltonian conform to theoretical expectations? 

We now introduce the observables computed in this work to address these questions.

\subsubsection{Susceptibility}
The analysis of the Replica Symmetric (RS) Hamiltonian within the framework of field theory~\cite{pimentel:02,parisi:13}, identifies three distinct masses: replicon, anomalous, and longitudinal, along with their corresponding propagators (correlation functions).

In the spin-glass phase, the replicon mode exhibits the highest degree of singularity. When the limit of the number of replicas $n$ in the analytical computation approaches zero, the anomalous and longitudinal modes converge and become indistinguishable at $h\neq0$. Consequently, the two fundamental propagators of the theory, $G_R(\boldsymbol{x} - \boldsymbol{y})$ and $G_A(\boldsymbol{x} - \boldsymbol{y})$, are defined as follows:
\begin{equation}\label{eq:GR}
G_R(\boldsymbol{x}-\boldsymbol{y})= \overline{ \langle s_{\boldsymbol{x}} s_{\boldsymbol{y}} \rangle^2} -2
 \overline{ \langle s_{\boldsymbol{x}} s_{\boldsymbol{y}} \rangle \langle s_{\boldsymbol{x}}\rangle \langle s_{\boldsymbol{y}} \rangle}
 +  \overline{ \langle s_{\boldsymbol{x}} \rangle ^2 \langle s_{\boldsymbol{y}} \rangle^2}\;
\end{equation}
and
\begin{equation}\label{eq:GA}
 G_{\text{A}}(\boldsymbol{x}-\boldsymbol{y})= \overline{ \langle s_{\boldsymbol{x}} s_{\boldsymbol{y}} \rangle^2} -4
 \overline{ \langle s_{\boldsymbol{x}} s_{\boldsymbol{y}} \rangle \langle s_{\boldsymbol{x}}\rangle \langle s_{\boldsymbol{y}} \rangle}
 +3  \overline{ \langle s_{\boldsymbol{x}} \rangle ^2 \langle s_{\boldsymbol{y}} \rangle^2}\;.
\end{equation}

These two fundamental propagators are computed from 4-replica estimators. The details on how to compute them can be found in Appendixes A and E of Ref.~\cite{aguilar:24b}.
Associated with each two-point correlation functions one can define two susceptibilities as 
\begin{equation}\label{suscep}
    \chi_{\alpha}=\widetilde{G}_{\alpha}(\boldsymbol{0})\; \;\;\; \alpha\in \{R,A\} \, ,
\end{equation}
where $\widetilde{G}(\boldsymbol{k})$ is the discrete Fourier transform of $G(\boldsymbol{x})$. In absence of an external magnetic field, it is trivial to show that $\chi_{R}=\chi_{A}=\chi_{L}$. When $h\ne 0$ instead, the replicon susceptibility $\chi_{\text{R}}$ diverges at criticallity while the anomalous and longitudinal modes remains massive~\cite{temesvari:02}. Hence, the replicon susceptibility becomes dominant in the spin-glass phase.

\subsubsection{Second-moment correlation length}
As a standard procedure to study the phase transition of the system, we will observe the behavior of the second-moment correlation length
\begin{equation}\label{xi2}
    \xi_2 = \frac{1}{2\sin(\pi/ L)}\left(\frac{\widetilde{G}_R(0)}{\widetilde{G}_R(\boldsymbol{k_1})}-1\right)^{1/2}\, ;
\end{equation}
where $\boldsymbol{k_1} = (2\pi/L,0,0,0,0)$ (or permutations). At a critical point, the quantity $\xi_2/L$ presents scale invariance and behaves as
\begin{equation} \label{eq:xi2_scale_invariance}
    \dfrac{\xi_2}{L} = f_\xi (L^{1/\nu}t) + L^{-\omega}g_\xi(L^{1/\nu}t) + \dots \, , 
\end{equation}
where $t=(T-\Tc)/\Tc$ is the reduced temperature and $\omega$ denotes the leading correction-to-scaling exponent. If a critical point indicating a phase transition exists, Eq.~\eqref{eq:xi2_scale_invariance} characterizes the behavior of our system in the vicinity of this point. Consequently, a plot of $\xi_2/L$ versus the temperature $T$, would reveal the intersection of the curves for different system sizes $L$ at the critical temperature $\Tc$.

Unfortunately, the second-moment correlation length does not exhibit the expected intersection for lower dimensions, as noted in Refs.~\cite{janus:12,leuzzi:09}. According to Refs.~\cite{leuzzi:09,janus:12,aguilar:24}, the anomalous behavior of the propagator at wave vector $\boldsymbol{k=0}$ introduces significant corrections to the leading scaling behavior described by Eq.~\eqref{eq:xi2_scale_invariance}.

To address this issue, we compute an alternative scale-invariant quantity, first introduced in Ref.~\cite{janus:12}: the dimensionless ratio of propagators with higher momenta, denoted as $R_{12}$

\begin{equation}\label{eq:R12}
    R_{12}=\frac{\widetilde{G}_R\left(\boldsymbol{k}_{1}\right)}{\widetilde{G}_R\left(\boldsymbol{k}_{2}\right)}\;.
\end{equation}
Here, $\boldsymbol{k}_{1}$ and $\boldsymbol{k}_{2}$ are the smallest nonzero momenta compatible with periodic boundary conditions, namely
$\boldsymbol{k}_{1} = (2\pi/L,0,0,0,0)$ and $\boldsymbol{k}_{2}= (2\pi/L,\pm 2\pi/L,0,0,0)$ (and permutations). 

\subsubsection{Study of RS effective Hamiltonian}

Besides the standard study of the phase transition and universality class, we will also evaluate the validity of some predictions from the replicated field theory. 

The effective Hamiltonian describing the $d$-dimensional Ising spin-glass in a magnetic field is complex, with three masses and eight cubic couplings. A detailed derivation starting from the microscopic Hamiltonian (Eq. \eqref{hamiltonian}) can be found in Ref.~\cite{temesvari:02b}. To study this theory, one introduces the following magnitude
\begin{equation}\label{eq:lambda_r}
    \lambda_r= \frac{w_{2,r}}{w_{1,r}}
\end{equation}
which represents the ratio between the renormalization vertices of the theory, $w_{1,r}$ and $w_{2,r}$. It can be shown (see Ref.\cite{parisi:13} and Appendix C of Ref.\cite{aguilar:24b}) that $\lambda_r$ can be computed as 
\begin{equation}\label{eq:lambda_r_computed}
    \lambda_r=\frac{ \omega_2}{ \omega_1}\;,
\end{equation}
where $ \omega_1$ and $ \omega_2$ are the following three-point connected correlation functions at zero external momentum:
\begin{equation}\label{eq:def-omega1}
    \omega_1= \frac{1}{V}\sum_{\boldsymbol{x} \boldsymbol{y} \boldsymbol{z}} \overline{\langle s_{\boldsymbol{x}} s_{\boldsymbol{y}}\rangle_c \langle s_{\boldsymbol{y}} s_{\boldsymbol{z}}\rangle_c \langle s_{\boldsymbol{z}} s_{\boldsymbol{x}}\rangle_c}\;,
\end{equation}
\
\begin{equation}\label{eq:def-omega2}
    \omega_2= \frac{1}{2V}\sum_{\boldsymbol{x}\boldsymbol{y}\boldsymbol{z}}\overline{\langle s_{\boldsymbol{x}} s_{\boldsymbol{y}} 
    s_{\boldsymbol{z}}\rangle^2_c}\;,
\end{equation}
where the subscript $c$ denotes \emph{connected} correlation functions. 

To compute these correlation functions, six real replicas are required. However, one can also compute the three- and four-replica estimators of $\omega_1$ and $ \omega_2$ (see Appendix C of Ref.\cite{aguilar:24b}), which in general differ from the six-replica value, but according to the Replica Symmetric field theory, coincide with it in the vicinity of $T_c$ \cite{parisi:13}. 

We will verify this theoretical prediction by computing the three estimators for a system in the presence of a magnetic field. In the absence of a field, $w_{2,r}=0$, and the theory simplifies. Furthermore, we will also check the value of $\lambda_r(T_c^+)$, defined as 
\begin{equation}
     \lambda_r(T_c^{+}) = \lim_{T\to T_c^+}\lim_{L\to\infty} \lambda_r(L,T)\,,
\end{equation}
since a value $\lambda_r(T_c^+)\in [0,1]$ indicates a continuous phase transition, while $\lambda_r(T_c^+)>1$ would imply a quasi-first-order phase transition \cite{holler:20}.

\section{Finite Size Scaling}\label{sec:FSS}
The estimation of the critical exponents in the infinite-volume system has been done by using Finite Size Scaling (FSS) techniques \cite{Amit-Martin, ballesteros:00}. In particular, we are interested in computing the critical temperature $T_c$, the correlation length exponent $\nu$, the exponent $\eta$, and, when possible, the leading correction-to-scaling exponent $\omega$. 

The interested reader will find the details of the computation in ~\ref{app:FSS}. We now include in the main text, for the reader's convenience, the expressions we have used for the numerical computation of the quantities.

Let us suppose we consider two dimensionless quantities $f(L,t)$, and $g(L,t)$, which scale in the same way (see details in ~\ref{app:FSS}). We can obtain the correction-to-scaling exponent $\omega$, through the quotient method \cite{nightingale:76, ballesteros:96b}. 
\begin{equation}\label{eq:FSS_omega}
    Q(g) = \frac{g(sL,t^*_L)}{g(L,t^*_L)} = 1 + B ^{g, f}_s L^{-\omega}\;,
\end{equation}
being $t^*_L$ a given reduced critical temperature obtained at the crossing point of $f(L,t)$ for two sizes $L$ and $sL$, and $s$ a scaling factor.
From this expression, we can determine the correction-to-scaling exponent $\omega$, with a fit with only two free parameters: $\omega$ and the non-universal amplitude $ B^{g, f}_s$. In particular, we can compute $Q(\xi_2/L)$ at the crossing points of $R_{12}$, and $Q(R_{12})$ at the crossing points of $\xi_2/L$. 

We can also compute an effective, size-dependent, $\nu(L)$ as 
\begin{equation}\label{eq:quotient_nu_xi}
    \nu(L)= \Bigg(\frac{\log(Q(\partial_\beta \xi_2))}{\log(s)}-1\Bigg)^{-1}
\end{equation}
which then must be extrapolated to the infinite volume value $\nu$ as 
\begin{equation}\label{eq:extrapola_nu}
    \nu(L)-\nu=AL^{-\omega}\;.
\end{equation}
So, once we have computed $\omega$ via Eq.~\eqref{eq:FSS_omega}, we can obtain $\nu$ through a fit with just two free parameters. Equivalently, $\eta$ can be computed from the dimensional spin-glass susceptibility $\chi_{\mathrm{SG}}$ remembering that $\chi_{SG} \sim |t|^{-\gamma}$ with $\gamma = (2-\eta)\nu$, so that 
\begin{equation}\label{eq:quotient_eta_chi}
    Q(\chi_{SG}) = \frac{\chi_{\mathrm{SG}}(sL,t^*_L)}{\chi_{\mathrm{SG}}(L,t^*_L)}= s^{2-\eta} + \mathcal{O}(L^{-\omega})
\end{equation}
and
\begin{equation}\label{eq:extrapola_eta}
    \eta(L) = 2- \frac{\log(Q(\chi_{\mathrm{SG}}))}{\log(s)} = \eta + BL^{-\omega}\;,
\end{equation}
where $B$ is a suitable constant.

\section{Numerical simulations}\label{sect:simulations}
We have conducted Monte Carlo simulations of the system defined by the Hamiltonian appearing in Eq.~\eqref{hamiltonian} for two values of the magnetic field: $h=0$ and $h=0.075$. We use modern CPUs capable of executing atomic instructions over 128-bit registers, and, since our system can be easily encoded in Boolean variables, we can follow a multi-spin coding approach that enables us to simulate 128 samples of the system in parallel. 

To ease the thermalization process, we have implemented the Parallel Tempering (PT) algorithm \cite{swendsen:87,hukushima:96}. We follow a very strict thermalization protocol, based on Ref.~\cite{billoire:18} which ensures that every single sample is thermalized before measuring any observable. 

The values of the lattice sizes ($L$) simulated, the number of samples for each lattice size, as well as other relevant simulation parameters for the system with zero and non-zero magnetic field can be consulted in Tables \ref{tabla_beta_c_H0} and \ref{tabla_beta_c_H} respectively. 

More details on the implementation of the Metropolis Algorithm on multi-spin coding simulations, the thermalization protocol and the measurement of the propagators in a multi-spin coding approach can be consulted in Appendices D and E of Ref.~\cite{aguilar:24b}. 
\begin{table}[h]
\centering
\begin{tabular}{r r c r r} \hline \hline
~~~$L$ & ~~~$N_{\mathrm{samp}}$ & ~~~$N_{T}$ & ~~~$T_{\text{min}}$ & ~~~$T_{\text{max}}$ \\ \hline
4 & 25600 & 10 & 2.3 & 3\\
5 & 25600 & 12 & 2.3 & 3\\
6 & 25600 & 14 & 2.3 & 2.95\\
7 & 25600 & 23 & 2.3 & 3\\
8 & 6400& 29 &2.3 & 3  \\
9 & 6400 & 32 & 2.3 & 3\\
10 & 6400 & 36 & 2.3 & 3\\ 
11 & 3200& 44& 2.3 & 3\\
12& 1280& 58& 2.3 & 3.041\\
14& 1280 & 46 & 2.54 & 3\\
\hline \hline
\end{tabular}
\caption{Parameters of the simulations for the five-dimensional Edwards-Anderson model at zero external filed $h=0$. $N_{\mathrm{samp}}$ is the number of samples simulated. $N_{T}$ is the number of temperatures simulated, which are uniformly distributed between $T_{\text{min}}$ and $T_{\text{max}}$.}
\label{tabla_beta_c_H0}
\end{table}

\begin{table}[h]
\centering
\begin{tabular}{r r c r r} \hline \hline
~~~$L$ & ~~~$N_{\mathrm{samp}}$ & ~~~$N_{T}$ & ~~~$T_{\text{min}}$ & ~~~$T_{\text{max}}$ \\ \hline
6 & 25600 & 13 & 2.3 & 2.95\\
7 & 25600 & 16 & 2.3 & 3\\
8 & 12800& 22 &2.3 & 3  \\
9 & 10240 & 28 & 2.3 & 3\\
10 & 10240 & 36 & 2.3 & 3\\ 
\hline \hline
\end{tabular}
\caption{Parameters of the simulations for the five-dimensional Edwards-Anderson model for nonzero magnetic field $h=0.075$. $N_{\mathrm{samp}}$ is the number of samples simulated. $N_{T}$ is the number of temperatures simulated, which are uniformly distributed between $T_{\text{min}}$ and $T_{\text{max}}$.}
\label{tabla_beta_c_H}
\end{table}

\section{Phase transition at $h=0$}\label{sect:zero-h}
In this section, we will compute the critical temperature and the critical exponents in absence of an external magnetic field at Subsec.~\ref{subsec:critical_h0}. Then, we will provide an estimation of the lower critical dimension at Subsec.~\ref{subsec:lcd}. 
\subsection{Critical temperature and critical exponents}\label{subsec:critical_h0}
\begin{figure}
    \centering
    \includegraphics[width=0.8\textwidth]{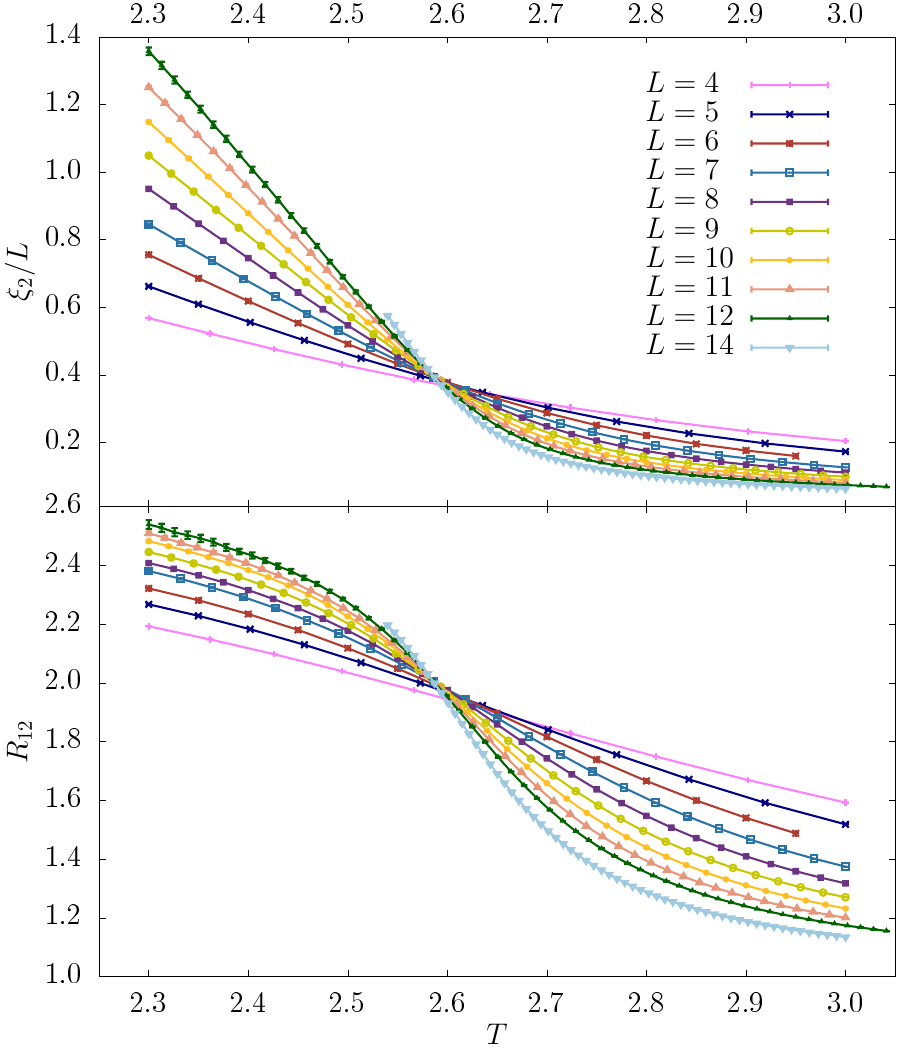}
    \caption{Second moment correlation length $\xi_2$, see Eq.~\eqref{xi2}, measured in units of the lattice size $L$ (top) and the dimensionless quotient $R_{12}$, see Eq.~\eqref{eq:R12}, (bottom), as a function of temperature $T$ for lattice sizes ranging from $L=6$ to $L=12$ and zero external magnetic field $h=0$.}
    \label{fig:cortesH0}
\end{figure}

We begin studying the phase transition in the absence of an external magnetic field. As a standard test, we observe the behavior of the dimensionless quantity $\xi_2/L$ as a function of the temperature $T$. Similarly, we have also examined the behavior of the observable $R_{12}$. Clear evidence of a phase transition can be seen in the intersections of Fig.~\ref{fig:cortesH0}, from which we can obtain a series of estimates for the critical temperature, see Table \ref{Table:Tc_exp_H0}.

\begin{table}[h]
\centering
\begin{tabular}{cccccc} \hline \hline
$L_1$& $L_2$& $T_{c}^R$ & $T_{c}^{\xi}$ & $\eta^{\xi}$ & $\nu^{\xi}$ \\ \hline
4 & 8 & 2.6223(9) & 2.6062(12) & -0.112(4) & 0.697(4)\\
5 & 10 & 2.5978(9) & 2.5924(11) & -0.168(4) & 0.717(7)\\
6 & 12 & 2.5887(12)& 2.5850(16) & -0.183(7) &  0.716(16)\\
7 & 14& 2.5870(9)& 2.5845(8) & -0.196(6) & 0.72(4) \\
\hline
6 & 7 & 2.608(3)& 2.588(6)&$-0.16(2)$& $0.75(4)$   \\
7 & 8 & 2.594(3)& 2.591(5) &$-0.14(2)$ & $0.66(3)$ \\
8 & 9 & 2.585(5)& 2.595(6) &$-0.10(3)$&$0.73(6)$\\
9 & 10& 2.585(5)& 2.585(6) &$-0.16(3)$&$0.71(5)$ \\
10 & 11 & 2.584(5)& 2.593(8)& $-0.10(4)$ &$0.65(7)$ \\
11 & 12 & 2.587(5)& 2.58(13) & $-0.15(9)$ &$0.9(3)$ \\
\hline
6 & 8 & 2.601(2)& 2.589(3)& $-0.15(2)$& $0.706(13)$ \\
7 & 9 & 2.589(2) & 2.592(3) &$-0.12(2)$ & $0.69(2)$ \\
8 & 10 & 2.585(2)& 2.589(2) &$-0.13(2)$ & $0.74(4)$ \\
9 & 11 & 2.584(3)& 2.589(2)&$-0.12(2)$& $0.68(3)$ \\
10 & 12 & 2.585(3)& 2.586(5)& $-0.15(3)$ & $0.73(8)$\\
\hline \hline
\end{tabular}
\caption{Values of the effective critical temperature and critical exponents. The top part of the table corresponds to $L_2=2L_1$, the middle part corresponds to $L_2=L_1+1$ and the bottom part to $L_2=L_1+2$. The third and fourth columns show the crossing points for $\xi_2/L$ and $R_{12}$ for consecutive lattice sizes. The fifth and sixth columns are the estimates for the critical exponents via the quotient method obtained at the intersection points of $\xi_2/L$. }
\label{Table:Tc_exp_H0}
\end{table}

Initially, we focus on determining the correction-to-scaling exponent $\omega$ through the quotient $Q(R_{12})$ at the intersection points of $\xi_2/L$ and vice versa, following the procedure explained in Sec.~\ref{sec:FSS}. These values are presented in Table~\ref{Table:w_corr_to_scaling}.
\begin{table}
\centering
\begin{tabular}{cccc} \hline \hline
$L_1$& $L_2$& $Q(R_{12})$ where $Q(\xi_2/L)=1$; &  $Q(\xi_2/L)$ where $Q(R_{12})=1$\\ \hline
4 & 8 & 1.0122(10) & 0.959(3)\\
5 & 10 & 1.0056(10) &0.985(4)\\
6 & 12 & 1.005(2) & 0.992(7) \\
7 & 14& 1.002(2) & 0.999(8)  \\
\hline \hline
\end{tabular}
\caption{ $h=0$-case. Quotients of $R_{12}$ at the crossing points $\xi_2/L$ and vice versa.}
\label{Table:w_corr_to_scaling}
\end{table}

To determine $\omega$ we try three different fits to Eq.~\eqref{eq:FSS_omega}: a fit to the $Q(R_{12})$ data with two free parameters $A_R$ and $\omega$, a fit to the $Q(\xi/L)$ data with different two free parameters $A_\xi$ and $\omega_\xi$, and a joint fit to both data sets, with three free parameters $A_R$, $A_\xi$ and a shared $\omega$. 

\begin{figure}
    \centering
    \includegraphics[width=0.8\textwidth]{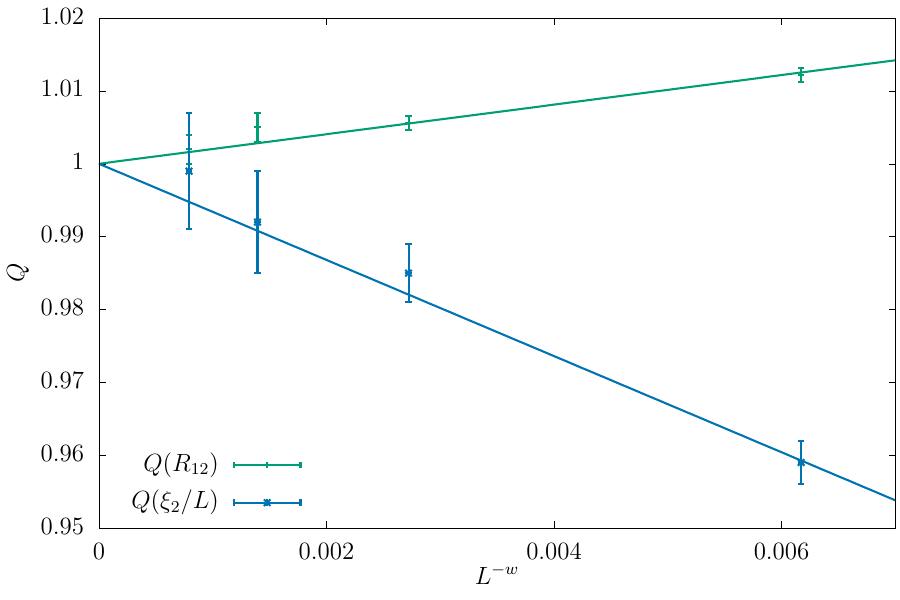}
    \caption{Quotient of $\xi/L$ at the crossing of $R_{12}$ and vice versa. The lines represent the joint fit to Eq.~\eqref{eq:FSS_omega} using both sets of data, which gives $\omega=3.7(4)$.}
    \label{fig:Q_corr_to_scaling}
\end{figure}

For the first fit we find $A_R=0.9(6)$, $\omega=3.1(4)$, with $\chi^2/\mathrm{d.o.f}=0.41$ and a $p$-value of $66\%$. For the second fit we obtain $A_\xi=-20(8)$, $\omega_\xi=4.5(3)$ with $\chi^2/\mathrm{d.o.f}=0.061$ and a $p$-value of $94\%$. Finally, a joint fit yields  $A_R=2(1)$, $A_\xi=-6(3)$, $\omega=3.7(4)$, $\chi^2/\mathrm{d.o.f}=0.41$, and a $p$-value of $83\%$ The numerical data, together with the joint fit can be seen in Fig.~\ref{fig:Q_corr_to_scaling}. 

Next, we compute the other two exponents needed to fully determine the universality class of the system. We begin by calculating the effective, size-dependent, critical exponents through the quotient method, see Eqs. \eqref{eq:quotient_nu_xi} and \eqref{eq:quotient_eta_chi}, at the intersection point of $\xi_2/L$. 

The values of the effective critical exponents $\nu$ and $\eta$ are shown in Table~\ref{Table:Tc_exp_H0}. The most accurate results, which we use for extrapolation, are obtained from considering $s=2$ (i.e., $L_2=2L_1$) and are presented in the top block of Table~\ref{Table:Tc_exp_H0}. However, we also include the results from $L_2=L_1+1$ (middle block) and $L_2=L_1+2$ (bottom block) as these are the only data available when we study the $h>0$ transition. These latter approaches provide a rough estimate of the critical exponents.

From the estimates of the critical exponents shown in the top block of Table \ref{Table:Tc_exp_H0}, we obtain the infinite volume values $\eta$ and $\nu$ through a fit to Eqs.~\eqref{eq:extrapola_nu} and \eqref{eq:extrapola_eta}. Using the value of $\omega=3.7(4)$ obtained from the joint fit we get: $\eta=-0.208(4)[5]$\footnote{We use round brackets for the statistical error and square brackets for the systematic error. In this case the systematic error is induced by the uncertainty in the value of $\omega$.}, with an amplitude $A_\eta=16(1)$, $\chi^2/\mathrm{d.o.f}=0.25$ and $p$-value of $77\%$;  and $\nu=0.729(11)[2]$ with an amplitude $A_\nu=-5(2)$, $\chi^2/\mathrm{d.o.f}=0.11$ and 
$p$-value of $89\%$. 

The best theoretical estimates for the critical exponents in five dimensions come from the 15th-order high-temperature series expansion of Ref~\cite{klein:91}, which yields the following central values: $\nu=0.726$ and $\eta=-0.38$. The big disagreement in the value of $\eta$ persist for other values of the spatial dimension, which are also considered in the cited reference. In $d=4$ the authors give the values $\nu=0.95$, $\eta=-0.11$ and in $d=3$ they obtain $\nu= 1.37$ and $\eta=0.03$. This values can be compared with the ones obtained by Monte Carlo simulations, see Table \ref{tab:exponentes_vs_dim}. The values of $\eta$ obtained in Ref.~\cite{klein:91} are not only in disagreement with the numerical ones, but grow in the opposite direction when changing $d$, and show an strange non-monotonic behavior given the fact that $\eta=0$ in six dimensions. Note, however, that the theoretical value of $\nu$ in five dimensions is in very good agreement with our numerical estimation. 

In addition, the $\epsilon$-expansion in $\epsilon=6-d$ was originally performed by Harris, Lubensky, and Chen in Ref.~\cite{harris:74} up to order $\epsilon$ and then expanded to order $\epsilon^3$ by Green in Ref.~\cite{green:85}. However, the asymptotic series converges poorly for $\epsilon=1$ and fails to give good estimates of the critical exponents.

Finally, using the values obtained for $\omega=3.7(4)$ and $\nu=0.729(11)[2]$ we can extrapolate to an infinite system size of the critical temperature \footnote{We have verified that our estimate of $\beta_c^\infty$ is stable under the inclusion of a second correction term in the fitting form,
\(
\beta_c(L_1)=\beta_c^\infty + A\,L_1^{-\omega-1/\nu} + B\,L_1^{-2\omega-1/\nu}
\). The resulting estimates are fully consistent with those obtained from the one–correction fit.}

\begin{equation}\label{eq:extrapol_beta}
    \beta_c(L_1)=\beta_c^\infty + A L_1^{-\omega-1/\nu}\;.
\end{equation}
Using the data of Table~\ref{Table:Tc_exp_H0}, one can try several extrapolations for the inverse critical temperature, depending on the observable considered.

An extrapolation using the effective, size-dependent, critical temperatures from $R_{12}$ and $L_2=2L_1$, gives us $\beta_c=0.38677(8)[15]$ (or $T_c=2.5854(5)[10]$) with $A=-6.1(2)$, $\chi^2/\mathrm{d.o.f}=0.53$, and a $p$-value of $58\%$. Alternatively, an extrapolation using the data from $\xi_2/L$ gives us $\beta_c=0.38706(11)[6]$ (or $T_c=2.5835(7)[4]$) with $A=-3.8(3)$, $\chi^2/\mathrm{d.o.f}=1.6$, and a $p$-value of $18\%$. Finally, a joint fit with both sets of data and forcing a common value of $T_c$ gives us $\beta_c=0.38688(7)[13]$ (or $T_c=2.5847(5)[9]$), with $A_R=-6.3(2)$, $A_{\xi}=-3.5(2)$, $\chi^2/\mathrm{d.o.f}=1.6$, and a $p$-value of $14\%$. 

Our estimate of the critical temperature is compatible with, but more accurate that, the two previous estimates  $\beta_c = 0.3889(13)$  ---or $T_c=2.571(9)$--- from \cite{klein:91} and  $\beta_c=0.3925(35)$ ---or $T_c=2.55(3)$--- from \cite{daboul:04}.

\subsection{Lower critical dimension}\label{subsec:lcd}

The lower critical dimension, $d_L$, is defined as the minimum spatial dimension below which a finite-temperature spin-glass phase cannot exist. Early experimental and numerical works show that a phase transition to a spin glass phase is present (in the absence of a field) in three dimensions \cite{gunnarsson:91,palassini:99,ballesteros:00}. On the contrary, no phase transition was found in $d=2$ \cite{morgenstern:79}, implying that $2<d_L<3$.

Early arguments based on the droplet picture of the spin glass phase suggested a lower critical dimension of $d_L = 2$~\cite{fisher:86}. Later studies, which computed the stiffness exponent $\theta$, raised the possibility that $d_L$ might depend on the disorder distribution~\cite{hartmann:01b}. However, this was clarified in Ref.~\cite{amoruso:03}, which showed that $d_L$ is in fact independent of the type of disorder and is approximately $d_L \approx 2.5$. More recently, both theoretical and numerical studies have provided evidence that a stable spin glass phase at finite temperature exists only in dimensions greater than $5/2$~\cite{franz:94, boettcher:05,maiorano:18}.

In this section, we apply a different method for the determination of the lower critical dimension, based on the numerical analysis of the values of the critical exponents $\eta$ and $\nu$ computed in different numerical simulations in different dimensions $d\ge 3$ \cite{janus:13,banos:12}.

The Renormalization Group theory predicts that the critical exponents are some function of the dimension of the system. At the lower critical dimension $d_L$, we should find $1/\nu\to 0 $ and  $\eta-2+d\to0$. 

We use this information to our advantage to estimate the lower critical dimension of the system. We build the Table \ref{tab:exponentes_vs_dim} with the values of the critical exponents for different values of the dimension, including the ones computed in this work for $d=5$. The reader can find the value of the exponents for three, four, and six dimensions in the referenced works of the caption of the Table \ref{tab:exponentes_vs_dim}. 

With those values of the exponents, we try several fits for $1/\nu(d)$ and $\eta(d)-2+d$. We find that a linear function does not describe the behavior of the critical exponents. For $1/\nu(d)$, we choose to fit to a quadratic function $f^{\nu}_{\mathrm{fit}}(d)=2+b(6-d)+c(6-d)^2$ that gives $\chi^2/\mathrm{d.o.f}=0.0028$ and a $p$-value of $95\%$. The function $f^{\nu}_{\mathrm{fit}}$ is chosen to impose that at the upper critical dimension $d_U=6$ we recover the Gaussian value of $\nu=1/2$. For $\eta$, we prefer a third degree fit of $\eta(d)-2+d$ to $f^{\eta}_{\mathrm{fit}}(d)=4+b_1(6-d)+b_2(6-d)^2+b_3(6-d)^3$. Again, we choose the function so that for $d=6$ we recover $\eta(d)-2+d=4$, i.e., the MF value $\eta=0$ \footnote{Because we have numerical data in $d=3,4,5$ and impose the analytic constraint $\eta(6)=0$ at the upper critical dimension, a cubic fit contains three free parameters for three numerical points, leaving zero residual degrees of freedom. Consequently, no standard estimate of the fit uncertainty can be provided. The uncertainty in the crossing point $d_L$ is instead obtained through a  parametric bootstrap (Gaussian resampling): each data point is perturbed according to its numerical error, $10^3$ synthetic datasets are generated, and the standard deviation of the resulting $d_L$ values is taken as the error bar.}.

\begin{table}[h]
    \centering
    \begin{tabular}{c c c} \hline \hline
         $d$ & $\eta$ & $\nu$ \\ \hline
         3 &  -0.3900(36) & 2.562(42) \\
         \hline
         4  & -0.320(13) & 1.068(7) \\
         \hline
         5 & -0.208(4) & 0.729(11)\\
         \hline
         6 & 0 & 1/2 \\ \hline \hline
    \end{tabular}
    \caption{Values of the critical exponents $\eta$ and $\nu$ as a function of the dimensionality of the system for the Ising Spin Glass. The values of the exponents in three dimensions were obtained by the Janus Collaboration in Ref. \cite{janus:13}. The four dimensional exponents are from Ref. \cite{banos:12}. In six dimensions, the system is at its upper critical dimension $d_U$, and the phase transitions is characterized by the Gaussian critical exponents $\eta=0$ and $\nu=1/2$, with their corresponding logarithmic corrections, which were computed in Ref. \cite{ruiz-lorenzo:17}. }
    \label{tab:exponentes_vs_dim}
\end{table}
The intersection of these functions with zero gives us two estimates for the lower critical dimension $d_L$. The best estimation of the critical dimension is provided by the fit to the $\eta$-data
\begin{equation}
     d_L^{\eta}=2.43(3) \, ,
\end{equation}
which is at about two standard deviation from the prediction $d_L=5/2$. 
Unfortunately, the $\nu$ data are  noisier than those of $\eta$, which strongly affects the determination of $d_L$ based on $\nu$. From the $\nu$ data, we obtain the estimate $d_L^{\nu}=1.9(2)$.
\begin{figure}
    \centering
    \includegraphics[width=0.8\textwidth]{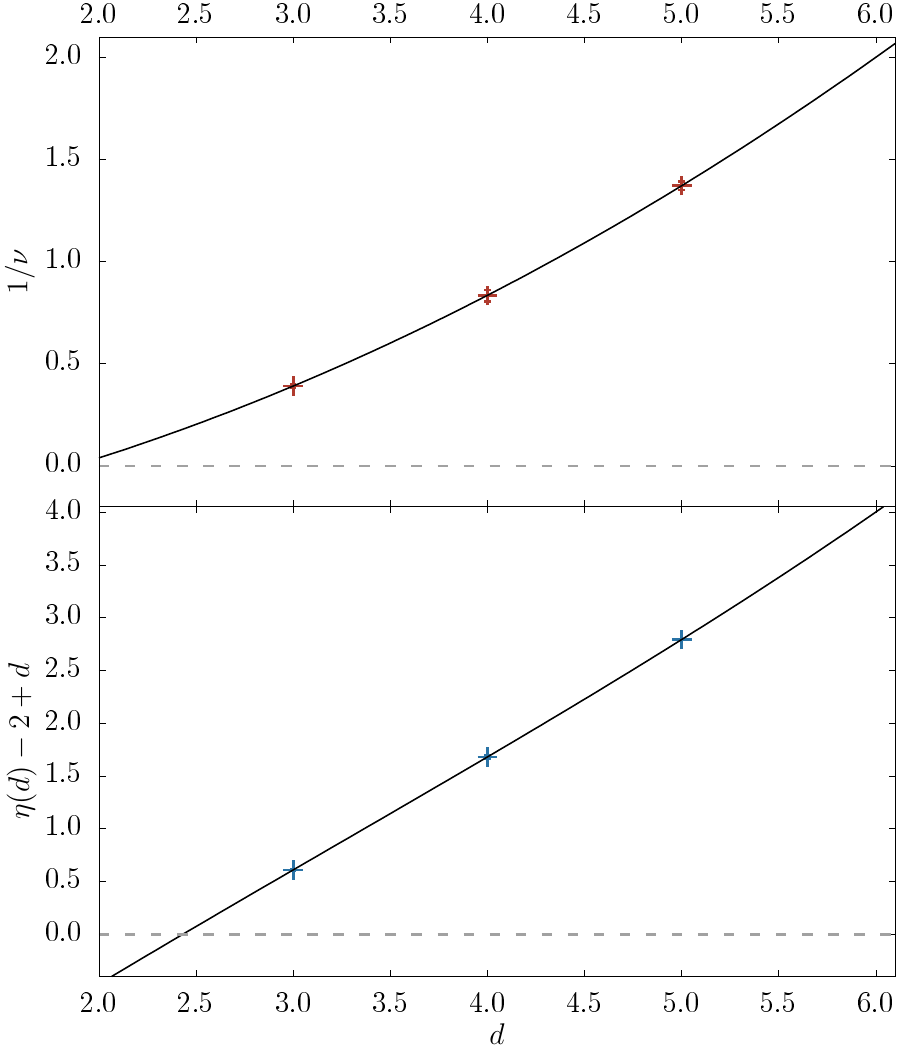}
    \caption{Critical exponents $\eta$ and $\nu$ as a function of the dimension for $h=0$. The points correspond to the value of the critical exponents that appear in Table \ref{tab:exponentes_vs_dim}. The solid lines correspond to the fits to the functions $f^\nu_{\mathrm{fit}}$ and $f^\eta_{\mathrm{fit}}$ defined in the text. The horizontal dashed lines at zero are used for reference to the eye. Statistical errors are present but, in most cases, smaller than the symbol size.}
    \label{fig:enter-label}
\end{figure}

\section{Phase transition in the presence of a field}\label{sect:not-zero-h}
In this section, we will present our results in the presence of an external magnetic field. We will start by discussing the replicon and anomalous susceptibilities in Subsec.~\ref{subsec:susceptibilities} to elucidate if we are deep enough in the spin-glass phase. Then, we will study the phase transition of the system for a non-zero external magnetic field and we will present our computations for the critical exponents in Subsec.~\ref{subsec:phasetransition_criticalexponents}. We continue by studying the extrapolation to the zero mode $\tilde G(\boldsymbol{0})$ in Subsec.~\ref{subsec:zeromode}, trying to explain the strong effect observed in the correlation length $\xi_2$. Finally, we compute and discuss the obtained results for the $\lambda_r$ parameter in Subsec~\ref{subsec:lambdar}.
\subsection{Replicon and anomalous susceptibilities}\label{subsec:susceptibilities}
Now we study the system in the presence of a magnetic field $h=0.075$. First, we address the issue of whether the magnitude of the field is big enough to ensure that our results are not a manifestation of the zero-field critical point. 

\begin{figure}
    \centering
    \includegraphics[width=0.8\textwidth]{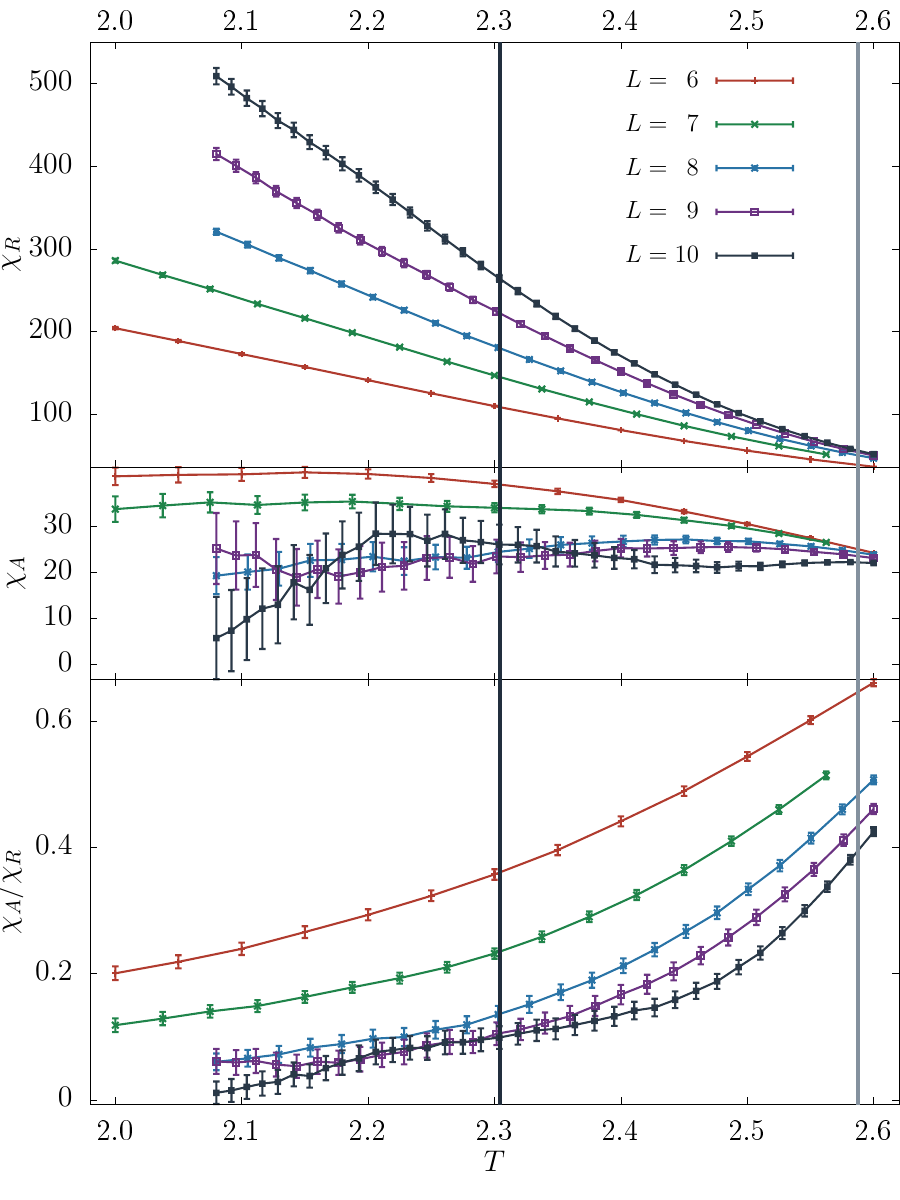}
    \caption{Replicon susceptibility $\chi_R$ (on the top), anomalous susceptibility (on the middle) and the quotient between the anomalous and replicon susceptibilities (on the bottom) as a function of temperature for different lattice sizes at $h=0.075$. The two vertical lines are our estimates for the critical temperature at $h=0$ (right grey line) and $h=0.075$ (left black line). See how $\chi_R$ grows rapidly with $L$ at low temperature, while $\chi_A/\chi_R$ goes to zero, ensuring that we are working far from the $h=0$ point of the de Almedia-Thouless line.}
    \label{fig:susceptibilities}
\end{figure}

In Fig.~\ref{fig:susceptibilities} we plot the replicon susceptibility $\chi_R$ (top panel), the anomalous susceptibility $\chi_A$ (middle panel), and its quotient $\chi_A/\chi_R$ (bottom panel). 

The replicated field theory predicts that both susceptibilities are identical in the zero-field case, but in the presence of a magnetic field, the replicon susceptibility becomes divergent in the spin glass phase, as we can observe in Fig.~\ref{fig:susceptibilities}.
\subsection{Phase transition and critical exponents}\label{subsec:phasetransition_criticalexponents}
Analogously to what we have done in the zero-field case, we now can study the phase transition. However, the presence of an external magnetic field causes some difficulties. On the first place,  the sample-to-sample fluctuations of the observables grow with respect to the case $h=0$, which forces us to simulate more samples to obtain comparable errors. The second difficulty is that the magnetic field makes it more computationally costly to reach equilibration.

As a consequence, we can only simulate system sizes up to $L=10$. This eliminates the possibility of performing such a detailed analysis as in the case $h=0$ where we have access to configurations of size $L$ and $2L$, so we will limit ourselves to studying the intersections of systems with $L$ and $L+1$, which, as we saw in the previous section, give us an approximate estimate of the value of the critical exponents.  

\begin{figure}
    \centering
    \includegraphics[width=0.8\textwidth]{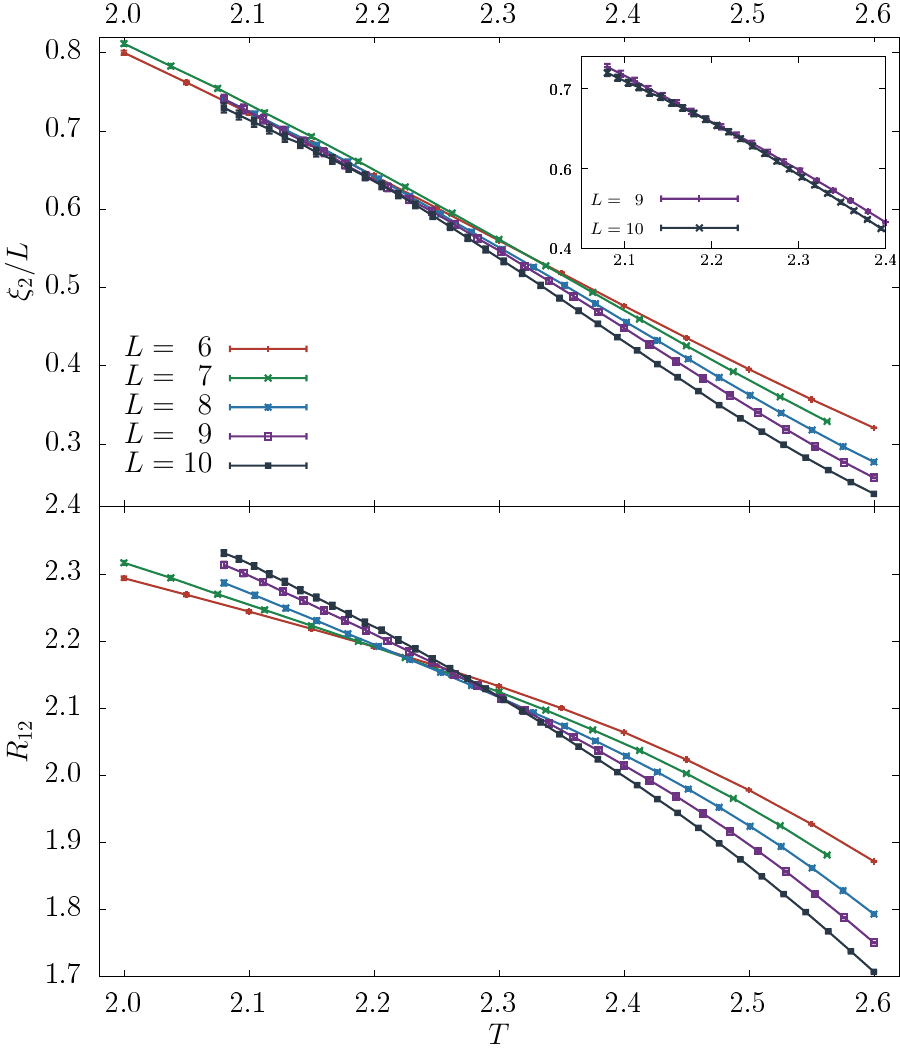}
    \caption{Second moment correlation length $\xi_2$, see Eq.~\eqref{xi2}, measured in units of the lattice size $L$ (on the top) and the dimensionless quotient $R_{12}$, Eq.~\eqref{eq:R12} (on the bottom), as a function of temperature $T$ for lattice sizes ranging from $L=6$ to $L=10$ and non-zero external magnetic field $h=0.075$.}
    \label{fig:cortesH75}
\end{figure}

The first important result of this section is the clear evidence of a phase transition shown by the crossing of curves for $R_{12}$ in Fig. \ref{fig:cortesH75}. The crossing values for systems of size $L$ and $L+1$ can be shown in Table \ref{Table:exponentesH75}.

The behavior of $\xi/L$ is, for the largest sizes simulated, compatible with a merging and not a crossing. This scenario is similar to the one found in four spatial dimensions~\cite{janus:12}, but different from the one we saw in the six-dimensional system \cite{aguilar:24b}, where we  found crossings for both $R_{12}$ and $\xi/L$.

These results have been previously attributed to the abnormal behavior of the zero-momentum mode of the propagator, $\widetilde G(\boldsymbol{0})$, caused by the negative tail of the probability distribution $P(q)$. This negative tail is a finite-size effect and should disappear in the infinite-volume limit. Some examples of $P(q)$'s in the presence of a magnetic field can be seen in Ref. \cite{aguilar:24b} for the six-dimensional system, and in Ref. \cite{aguilar:24} for the four-dimensional model. A possible alternative approach to study the phase transition in systems with a negative tail of $P(q)$ is developed in the next subsection \ref{sec:momentos}.

\begin{table}[h]
\centering
\begin{tabular}{ccccc} \hline \hline
$L_1$& $L_2$& $T_{c}^R$  & $\eta$ & $\nu$ \\ \hline
6 & 7 & 2.20(3) & 0.271(14)& 1.00(12)\\
7 & 8 & 2.23(3) & 0.189(15)& 0.75(8) \\ 
8 & 9 & 2.30(4) & 0.17(2)& 0.64(6) \\
9 & 10 & 2.30(3)  &0.10(2) & 0.75(17) \\
\hline 
6 & 8 & 2.207(6) & 0.231(16) & 0.84(4) \\
7 & 9 & 2.271(6) & 0.180(15) & 0.75(7) \\
8 & 10 & 2.300(5) & 0.13(2) & 0.69(6)\\
\hline \hline

\end{tabular}
\caption{Values of the effective critical temperature computed by the crossing points of $R_{12}$, $T_c^R$,  and effective critical exponents $\eta$ and $\nu$. The critical exponent $\eta$ have been through $\widetilde G(\boldsymbol{k_1})$ and $\nu$ have been computed via $\partial_{\beta}R_{12}$. }
\label{Table:exponentesH75}
\end{table}

The values for the effective critical temperature for a given pair of lattice sizes $(L_1, L_2)$ can be consulted in Table \ref{Table:exponentesH75}. The reader can observe how the value of $T_c(h=0.075)$ is significantly smaller than the zero-field critical temperature $T_c(h=0)=2.5847(5)$.

The computation of the critical exponents $\eta$ and $\nu$ has been done avoiding completely the anomalous $\boldsymbol{k}=0$ wave vector, i.e: we apply the quotient method at the intersection points of $R_{12}$ and extract $\eta$ from $\mathcal{F}=\widetilde G_R (\boldsymbol{k_1})$ and $\nu$ from $\partial_{\beta}R_{12}$, as explained in Sec. \ref{sec:FSS}. The results for the effective exponents can be seen in Table~\ref{Table:exponentesH75}. These exponents should be extrapolated to $L_1\to\infty$. However, given the limited range of lattice sizes and the size of our statistical errors, a controlled extrapolation is not reliable.  
We therefore restrict ourselves to reporting the effective exponents of Table~\ref{Table:exponentesH75}, which already show a mild systematic drift compatible with finite-size corrections.  
A global fit would require both a wider span of $L$ and an estimate of the correction-to-scaling exponent $\omega$.

In a recent paper~\cite{angelini:25}, an epsilon expansion in $\epsilon=8-d$ has been carried out to $\mathcal{O}(\epsilon)$ to compute the critical exponents of the phase transition to a spin glass phase in a magnetic field and at zero temperature. For $\epsilon=3$, they obtain the estimates $\nu=0.6875$, $\eta=-0.1875$, and $\omega=3$. We do not attempt a quantitative comparison with these values, since $\epsilon=3$ lies well outside the perturbative regime where the expansion is controlled.

\subsection{Extrapolation of the zero mode $\tilde G(\boldsymbol{0})$}\label{subsec:zeromode}
\label{sec:momentos}
\begin{figure}
    \centering
    \includegraphics[width=0.8\textwidth]{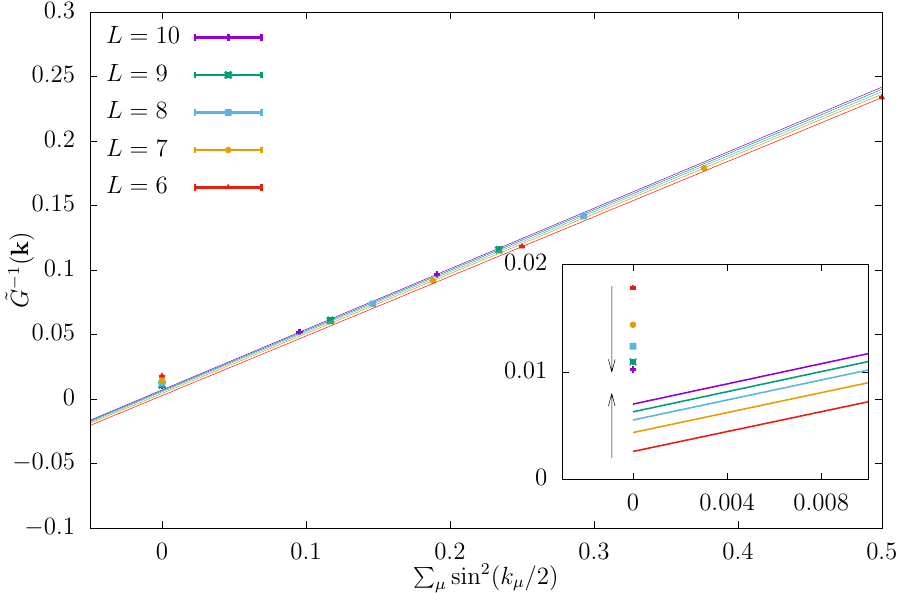}
    \caption{$\widetilde G^{-1}(\boldsymbol{k})$ vs $\sum_{\mu}\sin² (k_{\mu}/2)$ for $\beta=0.40$ and $L$ ranging from $L=6$ to $L=10$. The lines are a fit to the numerical data excluding the $\widetilde G(\boldsymbol{0})^{-1}$ point. Inset: Comparison between $\widetilde G(\boldsymbol{0})^{-1}$ and the extrapolated value $A(L,\beta)$.}
    \label{fig:momentos1}
\end{figure}

In this subsection, we follow the approach taken in Ref.~\cite{leuzzi:09}, and try to study in more detail the FSS in the $\boldsymbol{k}=\boldsymbol{0}$ mode of the propagator, which affects the values of $\chi_{SG}$ and $\xi_2$.  In Fig.~\ref{fig:momentos1} we plot $1/\tilde G_R(\boldsymbol{k})$ versus $\sum_\mu \sin^2(k_\mu/2)$, for $T=2.5$. This peculiar choice of variables is given by the behavior of the propagator for $T>T_c$, $L\to\infty$ and small wave number 
\begin{equation}
     \frac{1}{\widetilde G_R(\boldsymbol{k})} \simeq A + B \sum_{\mu} 4 \sin^2(k_{\mu}/2) \;\;,
\end{equation}
where $A=1/\chi_{\mathrm{SG}}$. 

In Fig.~\ref{fig:momentos1}, for each value of $L$ we plot three data points, corresponding to $\boldsymbol{k}=0$, $\boldsymbol{k}_1=(2\pi /L, 0,0,0,0)$ and permutations and $\boldsymbol{k}=(2\pi/L, \pm 2\pi/L,0,0,0)$ and permutations. The lines in Fig.~\ref{fig:momentos1} correspond to a linear fit excluding the $\boldsymbol{k}=0$ point. As one can see in the inset, the FSS for $\boldsymbol{k}=0$, and $|\boldsymbol{k}| > 0$ have opposite signs, which can have a strong effect on the behavior of $\xi_2$, which is a function of $\widetilde{G}_R(0)/\widetilde{G}_R(\boldsymbol{k_1})$. 

\begin{figure}
    \centering
    \includegraphics[width=0.8\textwidth]{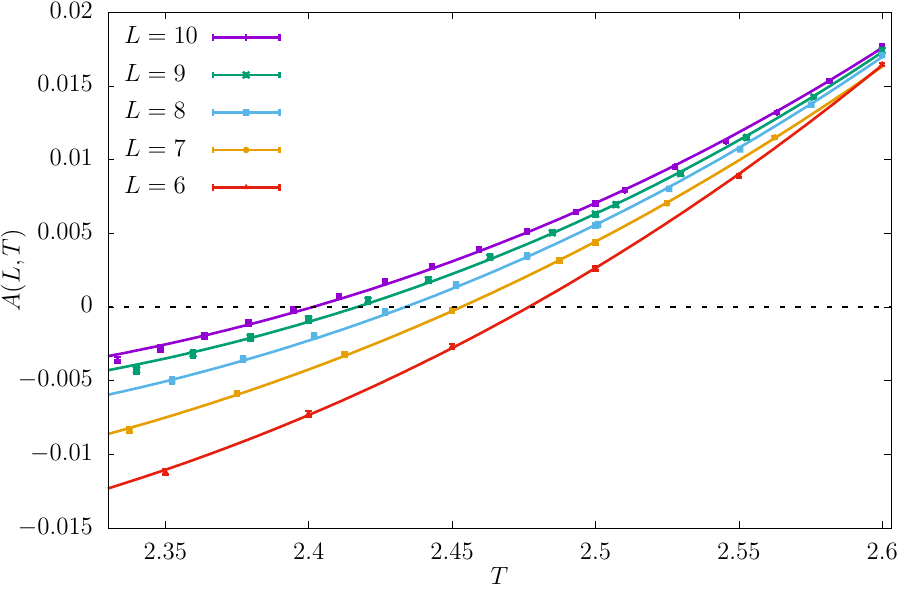}
    \caption{Plot of the extrapolated susceptibility $A(L,T)$ vs $T$ for all simulated lattice sizes. The curves are obtained by fitting the numerical data to a third degree polynomial. From the zeros of the fitted functions, we obtain the values for the effective critical temperature $T_c(L)$ that we show on Table~\ref{Table:beta_c_via_A}.}
    \label{fig:A(L)}
\end{figure}

We can give another estimate of $T_c$ using only data for $|\boldsymbol{k}|>0$. For each $T$ and each $L$, we can compute $A(L,T)$. These values can be seen in Fig.~\ref{fig:A(L)}. For each $L$, we fit $A(L,T)$ to a third degree polynomial,  and then obtain the effective critical temperature $T_c(L)$ by solving the equation $A(L,T_c(L))=0$. 

The values of $T_c(L)$ obtained by this procedure can be consulted in Table \ref{Table:beta_c_via_A}. Finally, we can estimate $T_c=\lim_{L\to\infty} T_c(L)$, by an extrapolation of the kind $\beta_c(L)=\beta_c+AL^{-1/\nu}$, obtaining $\beta_c=0.45(3)$ (or equivalently $T_c=2.22(15)$)and $\nu=0.65(55)$ ($\chi^2/\mathrm{d.o.f}=0.47$, and a $p$-value of 49\%). 

The values obtained from this extrapolation (in which we have excluded the $L=6$ data point) are compatible with those obtained in the previous section through the FSS analysis of $R_{12}$.   

\begin{table}[h]
\centering
\begin{tabular}{ccr} \hline \hline
$L$ & $T_c (L)$& $\beta_{c}(L)$  \\ \hline
6 &  2.4765(8)& 0.40379(14)\\
7 &  2.4529(9)&0.40768(15) \\ 
8 &  2.432(12)&0.4112(2) \\
9 &  2.414(17)&0.4143(3) \\
10&  2.398(12)&0.4170(2) \\\hline \hline
\end{tabular}
\caption{Values of the effective critical temperatures $T_c$ and it's inverse $\beta_c(L)$ computed from the zeros of  the fits to $A(L, T)$ shown in Fig.~\ref{fig:A(L)} for all lattice sizes. }
\label{Table:beta_c_via_A}
\end{table}

\subsection{$\lambda_r$ parameter}\label{subsec:lambdar}
Finally, we have studied the $\lambda_r$ parameter, defined by Eq. \eqref{eq:lambda_r_computed}. We have computed $\lambda_r$ using the three-, four-, and six-replica estimators of $\omega_1$ and $\omega_2$. 

This study will enable us to check the validity of the replica symmetric field theory for describing the critical behavior of Ising spin glasses in five dimensions. In particular, we have checked the prediction obtained in Ref.~\cite{parisi:13}, about the equivalence of the three-, four- and six-replica estimators of $\lambda_r$ at the critical temperature. 

Furthermore, the  different possible values of $\lambda_r$ have been linked with different kinds of phase transitions: $\lambda_r<1$ signals a continuous phase transition, while $\lambda_r>1$ would be related to a quasi first-order phase transition \cite{holler:20}.

\subsubsection{$\lambda_r$ computed with three, four, and six replicas}
\begin{figure}
    \centering
    \includegraphics[width=0.8\textwidth]{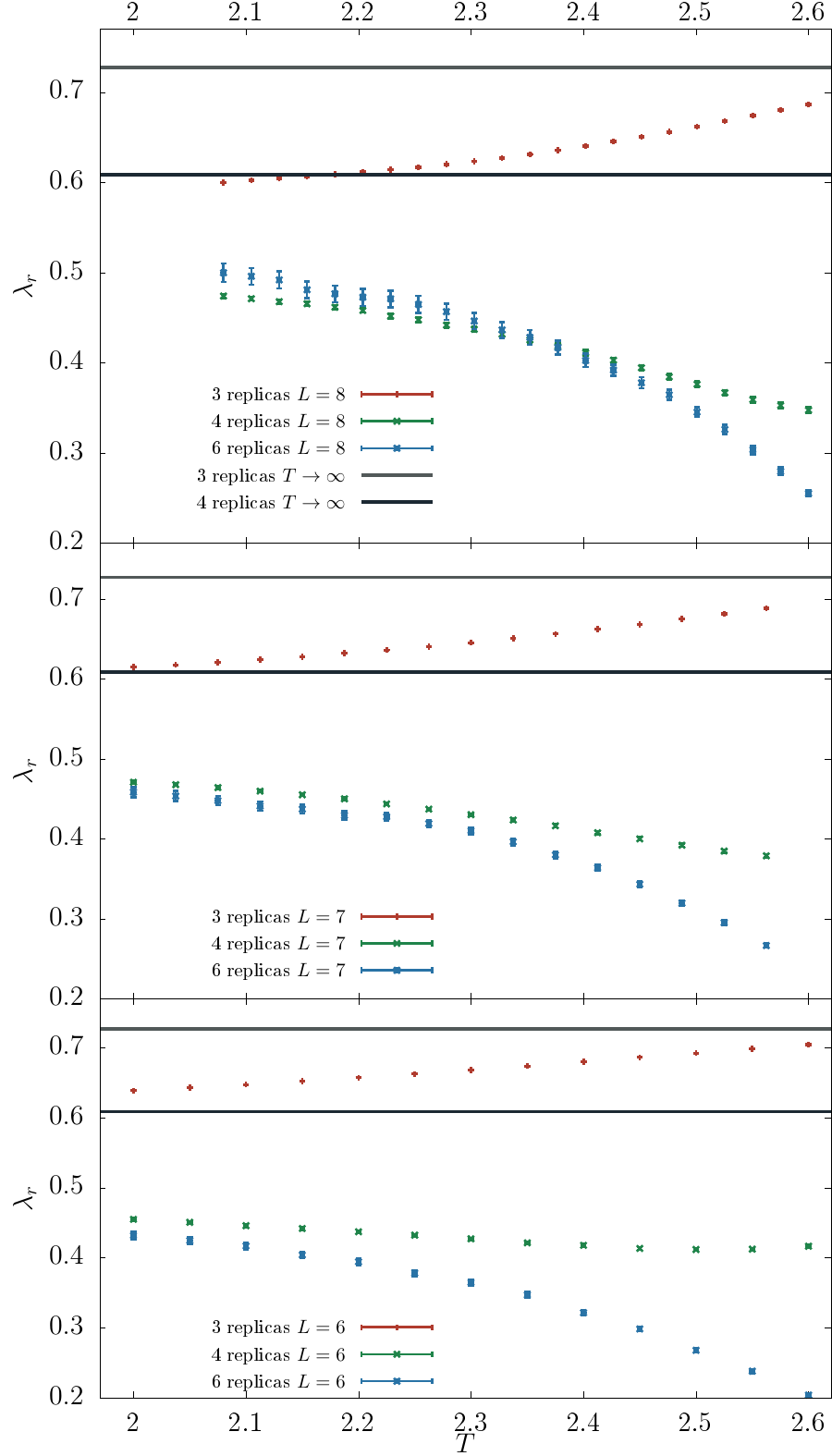}
    \caption{Three-, four-, and six-replica estimators od $\lambda_r$ as a function of the temperature for $L=6$ (top panel), $L=8$ (middle panel), and $L=10$ (bottom panel). The gray lines correspond to the $T\to\infty$ limit of the three-, and four-replica estimators.}
    \label{fig:lambda3-4-6}
\end{figure}

In Fig.~\ref{fig:lambda3-4-6} we show the three-, four-, and six-replica estimators of $\lambda_r$ for the different lattice sizes $L=6,8$ and 10. For reference, we also plot the infinite temperature values of the three- and four-replica estimators. 

The behavior found is very similar to that reported in the six-dimensional system of Ref.~\cite{aguilar:24b}. The three-replicas estimator seems to converge faster to its infinite temperature limit than the four-replica estimator. We also show that in the vicinity of $T_c(h=0.075)$, the four and six-replica estimators are compatible within one standard deviation. Also, note how the value of all three estimators converge with increasing $L$ to the same value near $T_c$, as predicted.

\subsubsection{Value of $\lambda_r(T_c^+)$}
\begin{figure}
    \centering
    \includegraphics[width=0.8\textwidth]{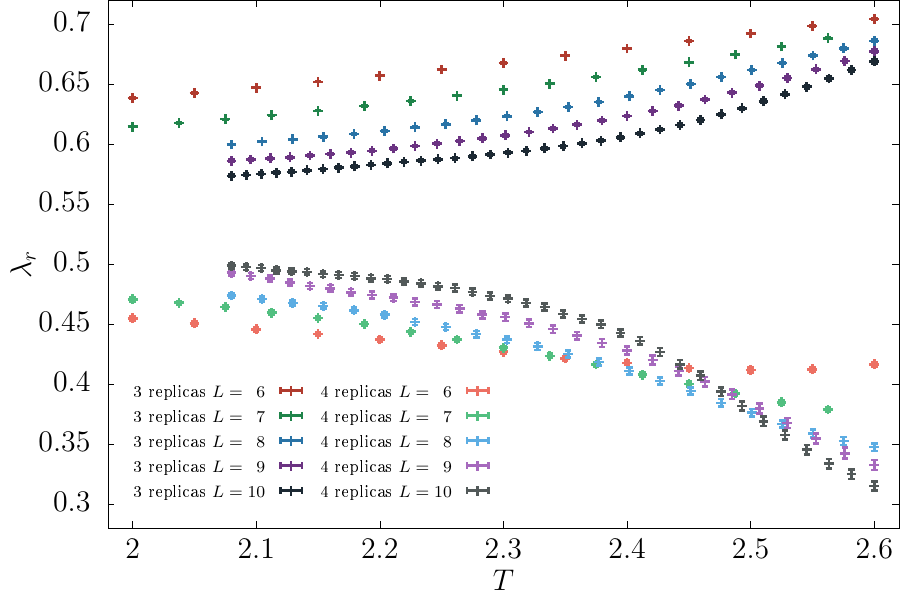}
    \caption{Three- and four-replica estimators of $\lambda_r$ as a function of the temperature. Note how the scaling corrections have opposite signs for each estimator. The four-replica estimator in compatible and less noisy than the six-replica one.}
    \label{fig:lambda_3-4}
\end{figure}

In Fig.~\ref{fig:lambda_3-4} we show the values of the three- and four-replica estimators of $\lambda_r$ for all our simulated lattice sizes. 

It is clear that the scaling corrections for each estimator have opposite sign, what enable us to give a safe estimate of $\lambda_r(T_c^+)=0.53(6)$. This is compatible with the prediction $\lambda_r\in [0,1]$, and the existence of a continuous phase transition. This also rules out the possibility of a quasi-first order transition, as proposed in Ref. \cite{holler:20} for $d\le6$.
\section{Conclusions}\label{sect:conclusions}

We have studied the critical behavior of the five-dimensional Edwards-Anderson model with and without an externally applied magnetic field.

For vanishing magnetic field, there is little doubt that we should find a phase transition in $d=5$, because
the lower critical dimension $d_\text{l}^{h=0}\approx 2.5$. Indeed, we obtain the critical exponents with high numerical accuracy. Our results for $d=5$ smoothly  interpolate with the previously known results for $d=3,4$~\cite{banos:12,janus:13} and $d=6$. Furthermore, a simple polynomial extrapolation for the anomalous dimension $\eta$
allows us to compute the lower critical dimension $d_\text{l}^{h=0}=2.43(3)$, in close agreement with a prediction
$d_\text{l}^{h=0}\approx 2.5$~\cite{franz:94, boettcher:05,maiorano:18}  that was obtained with entirely different methods featuring the low-temperature phase (rather than the critical temperature, as done here).

Instead, finding a de Almeida-Thouless line at $d=5$ 
could not be taken for granted from the outset. The 
Random-Field Ising model~\cite{nattermann:98} provides 
a good example (the externally applied magnetic field 
in a spin-glass is actually a random field, due to the 
gauge symmetry of spin glasses~\cite{toulouse:77}): 
while the lower critical dimension is 1 for the Ising 
model, it raises to $d_\text{l}^\text{RFIM}=2$ as soon 
as the random field is switched on. This abrupt change 
with external field has been emphasized by Leuzzi and 
Parisi~\cite{leuzzi:13}, in connection with the 
interpretation of Kac model simulations. Thus, it is well possible that $d_\text{l}^{h}>d_\text{l}^{h=0}$ for the Ising spin glass.

In spite of the above caveats, we do have found clear numerical evidence for a de Almeida-Thouless line in $d=5$. We have also been able to compute the critical exponents, including the leading corrections to scaling exponent. Of course, as in any other numerical investigation, one can never discard that qualitatively new behavior will eventually appear when still larger systems will be equilibrated (for more on this viewpoint, see Refs.~\cite{vedula:23,vedula:24,vedula:25}). Yet, out findings in $d=5$ are consistent with previous results obtained in four dimensions~\cite{janus:12},
as well as in the Kac model~\cite{leuzzi:09}.

The careful consideration of the zero-wave vector, $
\boldsymbol{p}=\boldsymbol{0}$ mode in the propagators 
has been crucial in our analysis (See also 
Ref.~\cite{janus:12}). The pathological behavior at
$\boldsymbol{p}=\boldsymbol{0}$ has been bypassed in 
two different ways: (i) we have considered a scale 
invariant quantity $R_{1,2}$ that is blind to $
\boldsymbol{p}=\boldsymbol{0}$ and (ii) we have 
extrapolated to $\boldsymbol{p}=\boldsymbol{0}$ our 
data for non-vanishing wave vectors. Both approaches 
yield consistent results. Interestingly enough, a 
previous study $d=6$~\cite{aguilar:24b} did not need to 
exclude $\boldsymbol{p}=\boldsymbol{0}$ from the 
analysis, although scaling corrections were certainly 
larger for that wavevector. Claryfiyng the physical 
reasons underlying this qualitative difference between 
$d=5$ and $d=6$ seems to us an important open problem.

Another important open question is the comparison of our estimates of the critical exponents in a field in $d=5$ with the results by other authors in different space dimensions (a similar comparison, but for vanishing magnetic field, was carried out in Sect.~\ref{subsec:lcd}). There are two natural candidates for this comparison,  namely the traditional two-loop computation in Refs.~\cite{charbonneau:17,charbonneau:19} and the Bethe lattice one-loop computation in Ref.~\cite{angelini:25}. The former calculation takes the form of a second-order polynomial in $(6-d)$ while the latter is a first order polynomial in $(8-d)$. Given the relatively large extrapolation needed to obtain prediction $d=5$, and the reduced accuracy of our Monte Carlo results in a field (as compared to those at $h=0$), we think that this comparison should be delayed until further work will have reduced the uncertainties.

\section{Acknowledgments}
We thank M. C. Angelini, S. Palazzi, G. Parisi, and T. Rizzo for interesting discussion and sharing the results of their analytical computation. We also thank Giorgio Parisi for discussions about our numerical results.
We acknowledge partial financial
support from Grant Nos. PID2024-156352NB-I00, PID2022-136374NB-C21, PID2020-112936GB-I00 funded by MCIN/AEI/10.13039/501100011033/FEDER, UE. and from Grant No.
GR24022 funded by Junta de Extremadura (Spain) and by
European Regional Development Fund (ERDF) “A way of
making Europe. We have performed the numerical simulations in the computing 
facilities of the Instituto de Computación Científica Avanzada (ICCAEx).

\clearpage
\appendix

\section{Computation details of Finite Size Scaling}\label{app:FSS}

In this Appendix, we will provide details of the computations of the quantities that characterize the phase transition, i.e. the critical temperature and the critical exponents, in an infinite system by extrapolating our finite-size simulations by using Finite Size Scaling (FSS) techniques \cite{Amit-Martin, ballesteros:00}. 

Our discussion of FSS will follow closely Ref.~\cite{banos:12}. A general dimensionless quantity $f(L,t)$ has the following FSS behavior
\begin{equation}
    f(L,t)=  F_0 (L^{1/\nu}t)+ L^{-\omega} F_1 (L^{1/\nu}t)
\end{equation}
where $t$ is the reduced temperature $t=\frac{T-T_c}{T_c}$. At large $L$ and small $t$, considering leading corrections, one obtains 
\begin{equation}\label{eq:scaling}
      f(L,t) =   F_0 (0) + L^{1/\nu} t F^\prime (0)+ L^{-\omega} F_1(0) + \dots
\end{equation}
Since dimensionless quantities are scale invariant, curves for lattice sizes $L$ and $sL$ ($s$ being a given scale factor) will share the same value at a given effective reduced critical temperature $t_L^*$, so 
\begin{equation}
\begin{split}
    &  F_0 (0) + L^{1/\nu} t^*_L  F^\prime (0)+ L^{-\omega} F_1(0)  =\\ &F_0 (0) + (sL)^{1/\nu} t^*_L  F^\prime (0)+ (sL)^{-\omega} F_1(0)
\end{split}
\end{equation}
which implies 
\begin{equation}
    t^*_L = \frac{T^*_L-T_c}{T_c}=A^f_s L^{-\omega-1/\nu}
\end{equation}
where the non-universal amplitude is 
\begin{equation}
    A^f_s  =\frac{(1-s^{-\omega})  F_1(0)}{(s^{1/\nu}-1) F_0^\prime(0) }\;.
\end{equation}

Now let's suppose we consider a second dimensionless quantity $g(L,t)$, which scales in the same way as $f(L,t)$. We can obtain the correction-to-scaling exponent $\omega$, through the quotient method \cite{nightingale:76, ballesteros:96b}. If we compute the value of $g(L,t)$ at $t=t^*_L$, obtained from the crossing point of $f(L,t)$ for two sizes $L$ and $sL$, we will have 
\begin{equation}
    g(L,t^*_L)\simeq  G_0(0) + A^{g,f}_sL^{-\omega}, 
\end{equation}
with $A^{g,f}_s=A^f_s G_0(0)+ G_1(0)$. Then, we can consider the quotient of this magnitude for the two sizes $L$ and $sL$, obtaining 
\begin{equation}\label{eqapp:FSS_omega}
    Q(g) = \frac{g(sL,t^*_L)}{g(L,t^*_L)} = 1 + B ^{g, f}_s L^{-w}\;. 
\end{equation}
From this expression we can determine the correction-to-scaling exponent $\omega$, with a fit with only two free parameters: $\omega$ and the non-universal amplitude $ B ^{g, f}_s$. In particular, we can compute $Q(\xi_2/L)$ at the crossing points of $R_{12}$, and $Q(R_{12})$ at the crossing points of $\xi_2/L$. 

The two other critical exponents $\nu$ and $\eta$ can also be obtained trough the quotient method. The usual way to obtain $\nu$ is through the study of the dimensional quantity $\partial_\beta \xi_2$. It follows from Eq.~\eqref{eq:scaling} that $\partial_\beta\xi_2$ scale as 
\begin{equation}
    \partial_\beta \xi_2(L,t) \simeq L^{1+1/\nu}  F_{0,\xi}(0) + \dots
\end{equation}
so that
\begin{equation}
    Q(\partial_\beta \xi_2) = \frac{\partial_\beta \xi_2(sL, t^*_L) }{\partial_\beta \xi_2(L,t^*_L)} = s^{1+1/\nu} +\mathcal{O}(L^{-w})
\end{equation}
so for a given value of $L$, $sL$ and $t_L^*$, we can compute an effective, size-dependent, $\nu(L)$ as 
\begin{equation}\label{eqapp:quotient_nu_xi}
    \nu(L)= \Bigg(\frac{\log(Q(\partial_\beta \xi_2))}{\log(s)}-1\Bigg)^{-1}
\end{equation}
which then must be extrapolated to the infinite volume value $\nu$ as 
\begin{equation}\label{eqapp:extrapola_nu}
    \nu(L)-\nu=AL^{-w}\;.
\end{equation}
So, once we have computed $\omega$ via Eq.~\eqref{eq:FSS_omega}, we can obtain $\nu$ through a fit with just two free parameters. Equivalently, $\eta$ can be computed from the dimensional spin-glass susceptibility $\chi_{\mathrm{SG}}$ remembering that $\chi_{\mathrm{SG}\sim |t|^{-\gamma}}$ and $\gamma = (2-\eta)\nu$, so that 
\begin{equation}\label{eqapp:quotient_eta_chi}
    Q(\chi_{SG}) = \frac{\chi_{\mathrm{SG}}(sL,t^*_L)}{\chi_{\mathrm{SG}}(L,t^*_L)}= s^{2-\eta} + \mathcal{O}(L^{-w})
\end{equation}
so 
\begin{equation}\label{eqapp:extrapola_eta}
    \eta(L) = 2- \frac{\log(Q(\chi_{\mathrm{SG}}))}{\log(s)} = \eta + BL^{-w}\;.
\end{equation}
We will follow this procedure to compute $\nu$ and $\eta$ when we study the phase transition at the $h=0$ point of the dAT line. However, it has been reported in previous works~\cite{janus:12,leuzzi:09, aguilar:24b} that when $h>0$, the propagator at wave vector $\boldsymbol k=\boldsymbol 0$ behaves anomalously. 

This anomaly has been related to the well-known negative tail of the $\overline{P(q)}$, a finite size issue that disappears in the infinite volume limit. This issue, however, causes some strong  corrections to the leading scaling behavior of those observables which depend on $\widetilde G_R(\boldsymbol{0})$, such as $\chi_{\mathrm{SG}}$ or $\xi_2$. 

In consequence, in our study of the system in the presence of a non-zero magnetic field $h$, we will avoid completely those observables. The exponent $\nu$ can be computed from $R_{12}$, Eq.~\ref{eq:R12}, applying the quotient method as
\begin{equation}
    Q(R_{12})=\frac{R_{12}(sL,t^*_L)}{R_{12}(L,t^*_L)}= s^{1/\nu} + \mathcal{O}(L^{-\omega})\;,
\end{equation}
so that 
\begin{equation}
    \nu(L)=\Bigg(\frac{\log(Q(R_{12}))}{\log(s)}\Bigg)^{-1}\;.
\end{equation}
In a similar way, we can compute $\eta$ through $\widetilde G_R(\boldsymbol{k_1})$, which scales in the same way as $\chi_{SG}$, so 
\begin{equation}
    \eta(L)= 2-\frac{\log(Q(\widetilde G_R(\boldsymbol{k_1}))}{\log(s)}\;.
\end{equation}

\section*{References}
\bibliographystyle{iopart-num}
\bibliography{biblio}

\providecommand{\newblock}{}
\begin{thebibliography}{10}
\expandafter\ifx\csname url\endcsname\relax
  \def\url#1{{\tt #1}}\fi
\expandafter\ifx\csname urlprefix\endcsname\relax\def\urlprefix{URL }\fi
\providecommand{\eprint}[2][]{\url{#2}}

\bibitem{wilson:74}
Wilson K~G and Kogut J 1974 {\em Physics Reports\/} {\bf 12} 75 -- 199 ISSN 0370-1573 \urlprefix\url{http://www.sciencedirect.com/science/article/pii/0370157374900234}

\bibitem{parisi:88}
Parisi G 1988 {\em Statistical Field Theory\/} (Addison-Wesley)

\bibitem{amit:05}
Amit D~J and Mart\'{i}n-Mayor V 2005 {\em Field Theory, the Renormalization Group and Critical Phenomena\/} 3rd ed (Singapore: World Scientific) \urlprefix\url{http://www.worldscientific.com/worldscibooks/10.1142/5715}

\bibitem{mydosh:93}
Mydosh J~A 1993 {\em Spin Glasses: an Experimental Introduction\/} (London: Taylor and Francis)

\bibitem{young:98}
Young A~P 1998 {\em Spin Glasses and Random Fields\/} (Singapore: World Scientific)

\bibitem{charbonneau2023spin}
Charbonneau P, Marinari E, Parisi G, Ricci-tersenghi F, Sicuro G, Zamponi F and Mezard M 2023 {\em Spin Glass Theory and Far Beyond: Replica Symmetry Breaking after 40 Years\/} (World Scientific)

\bibitem{bray:80}
Bray A~J and Roberts S~A 1980 {\em J. Phys. C: Solid St. Phys.\/} C {\bf 13} 5405

\bibitem{pimentel:02}
Pimentel I~R, Temesv\'ari T and De~Dominicis C 2002 {\em Phys. Rev. B\/} {\bf 65}(22) 224420 \urlprefix\url{https://link.aps.org/doi/10.1103/PhysRevB.65.224420}

\bibitem{mcmillan:84}
McMillan W~L 1984 {\em J. Phys.\/} C: Solid State Phys. {\bf 17} 3179

\bibitem{fisher:86}
Fisher D~S and Huse D~A 1986 {\em Phys. Rev. Lett.\/} {\bf 56}(15) 1601 \urlprefix\url{http://link.aps.org/doi/10.1103/PhysRevLett.56.1601}

\bibitem{bray:87}
Bray A~J and Moore M~A 1987 Scaling theory of the ordered phase of spin glasses {\em Heidelberg Colloquium on Glassy Dynamics\/} ({\em Lecture Notes in Physics\/} no 275) ed van Hemmen J~L and Morgenstern I (Berlin: Springer)

\bibitem{fisher:88}
Fisher D~S and Huse D~A 1988 {\em Phys. Rev. B\/} {\bf 38}(1) 373--385 \urlprefix\url{https://link.aps.org/doi/10.1103/PhysRevB.38.373}

\bibitem{yeo:15}
Yeo J and Moore M~A 2015 {\em Phys. Rev. B\/} {\bf 91}(10) 104432 \urlprefix\url{https://link.aps.org/doi/10.1103/PhysRevB.91.104432}

\bibitem{bray:11}
Bray A~J and Moore M~A 2011 {\em Phys. Rev.\/} B {\bf 83} 224408 (\textit{Preprint} \eprint{arXiv:1102.1675})

\bibitem{parisi:12}
Parisi G and Temesv\'ari T 2012 {\em Nucl. Phys.\/} B {\bf 858} 293 (\textit{Preprint} \eprint{arXiv:1111.3313})

\bibitem{singh:17b}
Singh R~R~P and Young A~P 2017 {\em Phys. Rev. E\/} {\bf 96}(1) 012127 \urlprefix\url{https://link.aps.org/doi/10.1103/PhysRevE.96.012127}

\bibitem{charbonneau:17}
Charbonneau P and Yaida S 2017 {\em Phys. Rev. Lett.\/} {\bf 118}(21) 215701 \urlprefix\url{https://link.aps.org/doi/10.1103/PhysRevLett.118.215701}

\bibitem{charbonneau:19}
Charbonneau P, Hu Y, Raju A, Sethna J~P and Yaida S 2019 {\em Phys. Rev. E\/} {\bf 99}(2) 022132 \urlprefix\url{https://link.aps.org/doi/10.1103/PhysRevE.99.022132}

\bibitem{holler:20}
H\"oller J and Read N 2020 {\em Phys. Rev. E\/} {\bf 101}(4) 042114 \urlprefix\url{https://link.aps.org/doi/10.1103/PhysRevE.101.042114}

\bibitem{janus:12}
Ba\~{n}os R~A, Cruz A, Fernandez L~A, Gil-Narvion J~M, Gordillo-Guerrero A, Guidetti M, Iniguez D, Maiorano A, Marinari E, Mart\'{i}n-Mayor V, Monforte-Garcia J, Mu{\~n}oz~Sudupe A, Navarro D, Parisi G, Perez-Gaviro S, Ruiz-Lorenzo J~J, Schifano S~F, Seoane B, Tarancon A, Tellez P, Tripiccione R and Yllanes D {2012} {\em Proc. Natl. Acad. Sci. USA\/} {\bf {109}} 6452

\bibitem{janus:14b}
Baity-Jesi M, Ba\~{n}os R~A, Cruz A, Fernandez L~A, Gil-Narvion J~M, Gordillo-Guerrero A, Iniguez D, Maiorano A, F M, Marinari E, Mart\'{i}n-Mayor V, Monforte-Garcia J, Mu{\~n}oz~Sudupe A, Navarro D, Parisi G, Perez-Gaviro S, Pivanti M, Ricci-Tersenghi F, Ruiz-Lorenzo J~J, Schifano S~F, Seoane B, Tarancon A, Tripiccione R and Yllanes D {2014} {\em Phys. Rev. E\/} {\bf 89} 032140 (\textit{Preprint} \eprint{arXiv:1307.4998})

\bibitem{janus:14c}
Baity-Jesi M, Ba\~{n}os R~A, Cruz A, Fernandez L~A, Gil-Narvion J~M, Gordillo-Guerrero A, Iniguez D, Maiorano A, F M, Marinari E, Mart\'{i}n-Mayor V, Monforte-Garcia J, Mu{\~n}oz~Sudupe A, Navarro D, Parisi G, Perez-Gaviro S, Pivanti M, Ricci-Tersenghi F, Ruiz-Lorenzo J~J, Schifano S~F, Seoane B, Tarancon A, Tripiccione R and Yllanes D {2014} {\em J. Stat. Mech.\/} {\bf 2014} P05014 (\textit{Preprint} \eprint{arXiv:1403.2622})

\bibitem{katzgraber:05b}
Katzgraber H~G and Young A~P 2005 {\em Physical Review B\/} {\bf 72} 184416

\bibitem{katzgraber:09}
Katzgraber H~G, Larson D and Young A 2009 {\em Physical review letters\/} {\bf 102} 177205

\bibitem{vedula:23}
Vedula B, Moore M~A and Sharma A 2023 {\em Phys. Rev. E\/} {\bf 108}(1) 014116 \urlprefix\url{https://link.aps.org/doi/10.1103/PhysRevE.108.014116}

\bibitem{vedula:24}
Vedula B, Moore M~A and Sharma A 2024 {\em Phys. Rev. E\/} {\bf 110}(5) 054131 \urlprefix\url{https://link.aps.org/doi/10.1103/PhysRevE.110.054131}

\bibitem{vedula:25}
Vedula B, Moore M and Sharma A 2025 {\em Phys. Rev. E\/} {\bf 111}(3) 034102 \urlprefix\url{https://link.aps.org/doi/10.1103/PhysRevE.111.034102}

\bibitem{leuzzi:09}
Leuzzi L, Parisi G, Ricci-Tersenghi F and Ruiz-Lorenzo J~J 2009 {\em Phys. Rev. Lett.\/} {\bf 103} 267201 (\textit{Preprint} \eprint{arXiv:0811.3435})

\bibitem{dilucca:20}
Dilucca M, Leuzzi L, Parisi G, Ricci-Tersenghi F and Ruiz-Lorenzo J~J 2020 {\em Entropy\/} {\bf 22} 250

\bibitem{Angelini:22}
Angelini M~C, Lucibello C, Parisi G, Perrupato G, Ricci-Tersenghi F and Rizzo T 2022 {\em Phys. Rev. Lett.\/} {\bf 128}(7) 075702 \urlprefix\url{https://link.aps.org/doi/10.1103/PhysRevLett.128.075702}

\bibitem{angelini:25}
Angelini M~C, Palazzi S, Parisi G and Rizzo T 2025 {\em Proceedings of the National Academy of Sciences\/} {\bf 122} e2511882122 \urlprefix\url{https://www.pnas.org/doi/abs/10.1073/pnas.2511882122}

\bibitem{aguilar:24b}
Aguilar-Janita M, Martin-Mayor V, Moreno-Gordo J and Ruiz-Lorenzo J~J 2024 {\em Phys. Rev. E\/} {\bf 109}(5) 055302 \urlprefix\url{https://link.aps.org/doi/10.1103/PhysRevE.109.055302}

\bibitem{harris:76}
Harris A~B, Lubensky T~C and Chen J~H 1976 {\em Phys. Rev. Lett.\/} {\bf 36}(8) 415--418 \urlprefix\url{http://link.aps.org/doi/10.1103/PhysRevLett.36.415}

\bibitem{janus:13}
Baity-Jesi M, Ba\~{n}os R~A, Cruz A, Fernandez L~A, Gil-Narvion J~M, Gordillo-Guerrero A, Iniguez D, Maiorano A, Mantovani F, Marinari E, Mart\'{i}n-Mayor V, Monforte-Garcia J, Mu{\~n}oz~Sudupe A, Navarro D, Parisi G, Perez-Gaviro S, Pivanti M, Ricci-Tersenghi F, Ruiz-Lorenzo J~J, Schifano S~F, Seoane B, Tarancon A, Tripiccione R and Yllanes D (Janus Collaboration) {2013} {\em Phys. Rev. B\/} {\bf 88} 224416 (\textit{Preprint} \eprint{arXiv:1310.2910})

\bibitem{miyazaki:13}
Miyazaki R and Nishimori H 2013 {\em Phys. Rev. E\/} {\bf 87}(3) 032154 \urlprefix\url{https://link.aps.org/doi/10.1103/PhysRevE.87.032154}

\bibitem{bernaschi:24b}
Bernaschi M, {González-Adalid Pemartín} I, Martín-Mayor V and Parisi G 2024 {\em Nature\/} {\bf 631} 749–754

\bibitem{dealmeida:78}
de~Almeida J~R~L and Thouless D~J 1978 {\em J. Phys. A: Math. Gen.\/} A {\bf 11} 983 \urlprefix\url{http://stacks.iop.org/0305-4470/11/i=5/a=028}

\bibitem{fernandez:22}
Fernandez L~A, Gonzalez-Adalid~Pemartin I, Martin-Mayor V, Parisi G, Ricci-Tersenghi F, Rizzo T, Ruiz-Lorenzo J~J and Veca M 2022 {\em Phys. Rev. E\/} {\bf 105}(5) 054106 \urlprefix\url{https://link.aps.org/doi/10.1103/PhysRevE.105.054106}

\bibitem{franz:94}
Franz S, Parisi G and Virasoro M 1994 {\em J. Phys.\/} (France) {\bf 4} 1657

\bibitem{boettcher:05}
Boettcher S 2005 {\em Phys. Rev. Lett.\/} {\bf 95}(19) 197205 (\textit{Preprint} \eprint{arXiv:cond-mat/0508061}) \urlprefix\url{http://link.aps.org/doi/10.1103/PhysRevLett.95.197205}

\bibitem{maiorano:18}
Maiorano A and Parisi G 2018 {\em Proceedings of the National Academy of Sciences\/} {\bf 115} 5129--5134 ISSN 0027-8424 (\textit{Preprint} \eprint{https://www.pnas.org/content/115/20/5129.full.pdf}) \urlprefix\url{https://www.pnas.org/content/115/20/5129}

\bibitem{parisi:13}
Parisi G and Rizzo T 2013 {\em Phys. Rev. E\/} {\bf 87}(1) 012101 \urlprefix\url{https://link.aps.org/doi/10.1103/PhysRevE.87.012101}

\bibitem{temesvari:02}
Temesv{\'a}ri T and De~Dominicis C 2002 {\em Phys. Rev. Lett.\/} {\bf 89} 097204 (\textit{Preprint} \eprint{arXiv:cond-mat/0207512})

\bibitem{aguilar:24}
Aguilar-Janita M, Franz S, Martin-Mayor V, Moreno-Gordo J, Parisi G, Ricci-Tersenghi F and Ruiz-Lorenzo J~J 2024 {\em Proceedings of the National Academy of Sciences\/} {\bf 121} e2404973121

\bibitem{temesvari:02b}
Temesv{\'a}ri T and De~Dominicis C I~R~P 2002 {\em Eur. Phys. J. B\/} {\bf 25} 361--372 (\textit{Preprint} \eprint{arXiv:cond-mat/0202162})

\bibitem{Amit-Martin}
Amit D~J and Matin-Mayor V 2005 {\em Field theory, the renormalization group, and critical phenomena: graphs to computers\/} (World Scientific)

\bibitem{ballesteros:00}
Ballesteros H~G, Cruz A, Fernandez L~A, Mart{\'\i}n-Mayor V, Pech J, Ruiz-Lorenzo J~J, Tarancon A, Tellez P, Ullod C~L and Ungil C 2000 {\em Phys. Rev.\/} B {\bf 62} 14237--14245 (\textit{Preprint} \eprint{arXiv:cond-mat/0006211})

\bibitem{nightingale:76}
Nightingale M 1976 {\em Physica A: Statistical Mechanics and its Applications\/} {\bf 83} 561 -- 572 ISSN 0378-4371 \urlprefix\url{http://www.sciencedirect.com/science/article/pii/0378437175900217}

\bibitem{ballesteros:96b}
Ballesteros H~G, Fernandez L~A, Mart{\'\i}n-Mayor V and Mu{\~n}oz~Sudupe A 1996 {\em Phys. Lett.\/} B {\bf 387} 125

\bibitem{swendsen:87}
Swendsen R~H and Wang J~S 1987 {\em Phys. Rev. Lett.\/} {\bf 58}(2) 86--88 \urlprefix\url{https://link.aps.org/doi/10.1103/PhysRevLett.58.86}

\bibitem{hukushima:96}
Hukushima K and Nemoto K 1996 {\em J. Phys. Soc. Japan\/} {\bf 65} 1604 (\textit{Preprint} \eprint{arXiv:cond-mat/9512035})

\bibitem{billoire:18}
Billoire A, Fernandez L~A, Maiorano A, Marinari E, Martin-Mayor V, Moreno-Gordo J, Parisi G, Ricci-Tersenghi F and Ruiz-Lorenzo J~J 2018 {\em Journal of Statistical Mechanics: Theory and Experiment\/} {\bf 2018} 033302 \urlprefix\url{http://stacks.iop.org/1742-5468/2018/i=3/a=033302}

\bibitem{klein:91}
Klein L, Adler J, Aharony A, Harris A~B and Meir Y 1991 {\em Phys. Rev. B\/} {\bf 43}(13) 11249--11273 \urlprefix\url{https://link.aps.org/doi/10.1103/PhysRevB.43.11249}

\bibitem{harris:74}
Harris A~B 1974 {\em Journal of Physics C: Solid State Physics\/} {\bf 7} 1671 \urlprefix\url{http://stacks.iop.org/0022-3719/7/i=9/a=009}

\bibitem{green:85}
Green J~E 1985 {\em Journal of Physics A: Mathematical and General\/} {\bf 18} L43 \urlprefix\url{https://dx.doi.org/10.1088/0305-4470/18/1/008}

\bibitem{daboul:04}
Daboul D, Chang I and Aharony A 2004 {\em Eur. Phys. J. B\/} {\bf 41} 231

\bibitem{gunnarsson:91}
Gunnarsson K, Svedlindh P, Nordblad P, Lundgren L, Aruga H and Ito A 1991 {\em Phys. Rev.\/} B {\bf 43} 8199--8203

\bibitem{palassini:99}
Palassini M and Caracciolo S 1999 {\em Phys. Rev. Lett.\/} {\bf 82} 5128--5131 (\textit{Preprint} \eprint{arXiv:cond-mat/9904246})

\bibitem{morgenstern:79}
Morgenstern I and Binder K 1979 {\em Phys. Rev. Lett.\/} {\bf 43}(21) 1615--1618 \urlprefix\url{https://link.aps.org/doi/10.1103/PhysRevLett.43.1615}

\bibitem{hartmann:01b}
Hartmann A~K and Young A~P 2001 {\em Phys. Rev. B\/} {\bf 64}(18) 180404 \urlprefix\url{http://link.aps.org/doi/10.1103/PhysRevB.64.180404}

\bibitem{amoruso:03}
Amoruso C, Marinari E, Martin O~C and Pagnani A 2003 {\em Phys. Rev. Lett.\/} {\bf 91}(8) 087201 \urlprefix\url{http://link.aps.org/doi/10.1103/PhysRevLett.91.087201}

\bibitem{banos:12}
Ba\~nos R~A, Fernandez L~A, Martin-Mayor V and Young A~P 2012 {\em Phys. Rev. B\/} {\bf 86}(13) 134416 (\textit{Preprint} \eprint{arXiv:1207.7014}) \urlprefix\url{http://link.aps.org/doi/10.1103/PhysRevB.86.134416}

\bibitem{ruiz-lorenzo:17}
Lorenzo J~J~R 2017 {\em Condensed Matter Physics\/} {\bf 20} 13601: 1–10 \urlprefix\url{https://arxiv.org/abs/1702.05072}

\bibitem{nattermann:98}
Nattermann T 1998 Theory of the {R}andom {F}ield {I}sing {M}odel {\em Spin glasses and random fields\/} ed Young A~P (Singapore: World Scientific)

\bibitem{toulouse:77}
Toulouse G 1977 {\em Communications on Physics\/} {\bf 2} 115

\bibitem{leuzzi:13}
Leuzzi L and Parisi G 2013 {\em Phys. Rev. B\/} {\bf 88} 224204 (\textit{Preprint} \eprint{arXiv:1303.6333})

\end{thebibliography}

\end{document}